\documentclass[11pt]{article}

\usepackage{amssymb}
\usepackage{amsmath}

\usepackage[pdftex]{graphicx}
\DeclareGraphicsExtensions{.png,.jpg,.eps} 

\usepackage{cases}
\usepackage{color}
\usepackage[ansinew]{inputenc} 
\usepackage{float}

\usepackage{subfigure}
\usepackage{slashbox}
\usepackage{setspace}
\usepackage{natbib}

\newtheorem{Rem}{Remark}

\title{\bf Pricing Variable
Annuity Guarantees in a Local Volatility framework}

\author{Griselda Deelstra and Gr\'{e}gory Ray\'ee \\[3mm]
Department of Mathematics, Universit\'e Libre de Bruxelles,\\
Boulevard du Triomphe, CP 210, Brussels 1050, Belgium}

\begin{document}

\maketitle


\begin{abstract}

In this paper, we study the price of Variable Annuity Guarantees,
especially of Guaranteed Annuity Options (GAO) and Guaranteed
Minimum Income Benefit (GMIB), and this in the settings of a
derivative pricing model where the underlying spot (the fund) is
locally governed by a geometric Brownian motion with local
volatility, while interest rates follow a Hull-White one-factor
Gaussian model. Notwithstanding the fact that in this framework,
the local volatility depends on a particularly complicated
expectation where no closed-form expression exists and it is
neither directly related to European call prices or other liquid
products, we present in this contribution different methods to
calibrate the local volatility model. We further compare Variable
Annuity Guarantee prices obtained in three different settings,
namely the local volatility, the stochastic volatility and the
constant volatility models all combined with stochastic interest
rates and show that an appropriate volatility modelling is
important for these long-dated derivatives. More precisely, we
compare prices of GAO, GMIB Rider and barrier types GAO obtained
by using local volatility, stochastic volatility and constant
volatility models.

\end{abstract}


\section{Introduction}

Variable Annuities are insurance contracts that propose a
guaranteed return at retirement often higher than the current
market rate and therefore they have become a part of the
retirement plans of many people. Variable Annuity products are
generally based on an investment in a mutual fund composed of
stocks and bonds (see for example Gao \cite{GaoJin} and Pelsser
and Schrager \cite{Pelsser2}) and they offer  a range of options
to give minimum guarantees and protect against negative equity
movement. One of the most popular type of Variable Annuity
Guarantees in Japan and North America is the Guaranteed Minimum
Income Benefit (GMIB). At her retirement date, a GMIB policyholder
will have the right to choose between the fund value at that time
or (life) annuity payments based on the initial fund value at a
fixed guarantee rate. Similar products are available in Europe
under the name Guaranteed Annuity Options (GAO). Many authors have
already studied the pricing and hedging of GMIBs and GAOs assuming
a geometric Brownian motion and a constant volatility for the fund
value (see for example Boyle and Hardy \cite{Boyle01, Boyle},
Ballotta and Haberman \cite{Ballotta}, Pelsser \cite{Pelsser},
Biffis and Millossovich \cite{Biffis}, Marshall et al.
\cite{Hardy2}, Chu and Kwok
\cite{Chu_Kwok}).\\

GAO and GMIB can be considered as long-dated options since their
maturity is based on the retirement date. When pricing long-dated
derivatives, it is highly recommended that the pricing model used
to evaluate and hedge the products takes into account the
stochastic behavior of the interest rates as well as the
stochastic behavior of the fund. Furthermore, the volatility of
the fund can have a significant impact and should not be
neglected. It has been shown in \cite{Boyle} that the value of the
fund as well as the interest rates and the mortality assumptions
influence strongly the cost of these guarantees. Some authors
consider the evolution of mortality stochastic as well (see for
example \citep{Ballotta} and \cite{Biffis}). In
\cite{van_Haastrecht_GAO}, van Haastrecht et al. have studied the
impact of the volatility of the fund on the
price of GAO by using a stochastic volatility approach.\\

Another category of models able to fit the vanilla market implied
volatilities are local volatility models introduced by Derman and
Kani and Dupire in 1994 in resp. \cite{Derman} and \cite{Dupire}
and recently extended to a stochastic interest rate framework by,
among others, Atlan \cite{Atlan}, Piterbarg \cite{Piterbarg2005}
and Deelstra and Rayee \cite{Deelstra-Rayee}. The main advantage
of local volatility models is that the volatility is a
deterministic function of the equity spot and time which avoids
the problem of working in incomplete markets in comparison with
stochastic volatility models. Therefore local volatility models
are more appropriate for hedging strategies. The local volatility
function is expressed in terms of implied volatilities or market
call prices and the calibration is done on the whole implied
volatility surface directly. Consequently, local volatility models
usually capture more precisely the surface of
implied volatilities than stochastic volatility models.\\

Stochastic volatility models have the advantage that it is
possible to derive closed-form solutions for some European
derivatives. In \citep{van_Haastrecht_GAO}, van Haastrecht et al.
have derived closed-form formulae for GAO prices in the
Sch\"{o}bel and Zhu stochastic volatility model combined with Hull
and White stochastic interest rates. However, the GMIB Rider, one
of the popular products traded by insurance companies in North
America (see \cite{GMIB}) has a more complicated payoff than a
pure GAO and therefore there exists no closed-form solution for
the price of a GMIB Rider, even not in the Sch\"{o}bel and Zhu
stochastic volatility model. The only way to evaluate a GMIB Rider
is by using numerical approaches like for example Monte Carlo
simulations.\\

In this paper, we study the prices of GAO, GMIB Riders and barrier
type GAOs in the settings of a two-factor pricing model where the
equity (fund) is locally governed by a geometric Brownian motion
with a local volatility, while interest rates follow a Hull-White
one-factor Gaussian model. In this framework, the local volatility
expression contains an expectation for which no closed-form
expression exists and which is unfortunately not directly related
to European call prices or other liquid products. Its calculation
can be done by numerical integration methods or Monte Carlo
simulations. An alternative approach is to calibrate the local
volatility from stochastic volatility models by establishing links
between local and stochastic volatility. A last calibration
approach presented in this paper is by adjusting the tractable
local volatility surface coming from a deterministic interest rates framework.\\

Furthermore, we compare Variable Annuity Guarantee prices obtained
in three different settings, namely, the local volatility, the
stochastic volatility and the constant volatility models all in
the settings of stochastic interest rates. We show that using a
non constant volatility for the volatility of the equity fund
value can have significant impact on the value of these Variable
Annuity Guarantees and that the impact generated by a local
volatility model is not equivalent to the one generated by a
stochastic volatility model, even if both are
calibrated to the same market data.\\

This paper is organized as follows: Section 2 introduces the local
volatility model with stochastic interest rates we use in this
paper to price Variable Annuity Guarantees. In Section
\ref{section_Calibration}, we present different approaches for the
calibration of the local volatility function. In Subsection
\ref{section:the loc vol function}, we derive an analytical
expression for the local volatility. In Subsection \ref{Monte
Carlo approach}, we explain in detail the different steps for
applying  a Monte Carlo method in the calibration procedure. Next,
in Subsection \ref{section: calibration Comparison between local
volatility} we give a link between the local volatility function
derived in a two-factor local volatility model and the tractable
one coming from the simple one-factor Gaussian model. Finally, in
Subsection \ref{section : mimicking stoch vol}, a link is given
between the stochastic volatility model and the local volatility
model in a stochastic interest rates framework. In Section
\ref{section Variable Annuities}, we present the three types of
Variable Annuity Guarantees we study in this paper, namely, the
GAO, the GMIB Rider and barrier types GAO. In Subsection
\ref{section_GAO_def}, we present the GAO, then, in Subsection
\ref{subsection:GMIB_Rider}, we define a GMIB Rider and finally in
Subsection \ref{section:Barrier GAO}, we study two types of
barrier GAO. Section \ref{section_numerical results} is devoted to
numerical results. In Subsection \ref{section Calibration
results}, we present the calibration procedure for the Hull and
White parameters and the calibration of the local volatility with
respect to the vanilla market. In Subsection \ref{section:
numerical results Comparison between local volatility}, we compare
local volatility surfaces obtained in a stochastic and in a
constant interest rates framework. Subsection \ref{section GAO
results}, \ref{section:GMIB_results} and \ref{section: Barrier GAO
results} investigate how the local volatility model behaves when
pricing GAO, GMIB Rider and barrier types GAO (respectively) with
respect to the Sch\"{o}bel-Zhu Hull-White stochastic volatility
model and the Black-Scholes Hull-White model. Conclusions are
given in Section \ref{section Conclusion}.

\section{The local volatility model with stochastic interest rates \label{section:the model}}

In this paper we consider a two-factor model where the volatility
of the fund value $S$ is a deterministic function of both time and
the fund itself. This function is known as ``local volatility". In
this model, the fund value $S$ is governed by the following
dynamics

\begin{eqnarray}
dS(t) = (r(t)-q)  S(t)  dt + \sigma(t,S(t)) S(t) dW^Q_{S}(t),
\label{spot_dynamic}
\end{eqnarray}

\noindent where interest rates follow a Hull-White one-factor
Gaussian model \citep{HullandWhite} defined by the
Ornstein-Uhlenbeck processes.

\begin{eqnarray}
dr(t) = [\theta(t) - \alpha(t)  r(t) ]  dt + \sigma_{r}(t)
dW^Q_{r}(t),  \label{rd_HW_dynamic}
\end{eqnarray}

\noindent where $\theta(t), \sigma_{r}(t)$ and $\alpha(t)$ are
deterministic functions of time. Equations (\ref{spot_dynamic})
and (\ref{rd_HW_dynamic}) are expressed under the risk-neutral
measure $Q$. We have chosen the very popular Hull-White model
since it is a tractable nontrivial interest rate model, allowing
closed-form solutions for many derivatives
which is useful for the calibration.\\

We assume that the dynamics of the fund and the interest rates are
linked by the following correlation structure:

\begin{eqnarray}
E^{Q}[dW^Q_{S} dW^Q_{r}] = \rho_{Sr} dt.
\end{eqnarray}

In the following we will denote this model by LVHW since it
combines a local volatility model with a Hull-White one-factor
Gaussian model. When $\sigma(t,S(t))$ equals a constant, this
model reduces to the Black-Scholes Hull-White model, denoted by
BSHW.\\

\section{Calibration \label{section_Calibration}}

Before using a model to price any derivatives, practitioners are
used to calibrate it on the vanilla market. The calibration
consists of determining all parameters present in the different
stochastic processes which define the model in such a way that all
European option prices derived in the model are as consistent as
possible with the corresponding market ones. More precisely, they
need a model which, after calibration, is able to price vanilla
options such that the resulting implied volatilities match the
market-quoted ones. \\

The calibration procedure for the LVHW model can be decomposed in
three steps: (i) Parameters present in the Hull-White one-factor
dynamics for the interest rates, namely $\theta(t), \alpha(t)$ and
$\sigma_{r}(t)$, are chosen to match European swaptions. Methods
for doing so are well developed in the literature (see for example
\citep{Brigo-Mercurio}). (ii) The correlation coefficient
$\rho_{Sr}$ is estimated from historical data. (iii) After these
two steps, one has to find the local volatility function which is
consistent with the implied volatility surface.\\

\subsection{The local volatility function \label{section:the loc vol function}}

In \cite{Deelstra-Rayee} we derived the local volatility
expression associated to a three-factor model by differentiating
the expression of a European call price ($C(K,T)$) with respect to
its strike $K$ and its maturity $T$. Following the same idea we
can derive the local volatility expression associated to the
two-factor model presented in Section \ref{section:the model}.
This leads to,

\begin{eqnarray}
\sigma(T,K) = \sqrt{\frac{\frac{\partial C(K,T)}{\partial T} + q
C(K,T) - q K \frac{\partial C(K,T)}{\partial K} + K P(0,T)
\mathbf{E}^{Q_T}[ r(T)  \textbf{1}_{ \{S(T)>K\}} ] }{ \frac{1}{2}
K^{2} \frac{\partial^{2} C}{\partial K^{2}}}}, \label{locvol C}
\end{eqnarray}

\noindent where $Q_T$ is the $T$-forward measure.\\

This local volatility function is not easy to calibrate with
respect to the vanilla market since there is no immediate way to
link the expectation term ($\mathbf{E}^{Q_T}[ r(T) \textbf{1}_{
\{S(T)>K\}} ]$) with vanilla option prices or other liquid
products. However, we present in this section three different methods to calibrate it.\\

Note that when assuming constant interest rates $(r=r(0))$,
equation (\ref{locvol C}) reduces to the simple Dupire formula
(see \cite{Dupire}) corresponding to the one-factor Gaussian case:

\begin{eqnarray}
\sigma(T,K) = \sqrt{\frac{ \frac{\partial C(K,T)}{\partial T} + (r
- q) K \frac{\partial C(K,T)}{\partial K}  + q
C(K,T)}{\frac{1}{2}K^2 \frac{\partial^{2} C(K,T)}{\partial
K^{2}}}}. \label{detreministic locvol C}
\end{eqnarray}

Since the market often quotes options in terms of implied
volatilities $\sigma_{imp}$ instead of option prices, it is more
convenient to express the local volatility in terms of implied
volatilities. The implied volatility of an option with price
$C(K,T)$, is defined through the Black-Scholes formula
($C^{mkt}(K,T) = C^{BS}(K,T,\sigma_{imp})$) and therefore
computing the derivatives of call prices through the chain rule
and substituting in equation (\ref{detreministic locvol C}) leads
to the following equation (see \citep{Wilmott2}),

\begin{eqnarray}
\sigma(T,K) &=&  \sqrt{\frac{ Vega^{BS} (\frac{\sigma_{imp}}{2T}+
 \frac{\partial \sigma_{imp}}{\partial T} + (r - q) K
\frac{\partial \sigma_{imp}}{\partial K} )}{\frac{1}{2} K^2
Vega^{BS} (\frac{1}{\sigma_{imp} K^2 T}+ \frac{2 d_+}{\sigma_{imp}
K \sqrt{T}} \frac{\partial \sigma_{imp}}{\partial K} +
\frac{\partial^2 \sigma_{imp}}{\partial K^2} + \frac{d_+
d_-}{\sigma_{imp}} (\frac{\partial \sigma_{imp}}{\partial K})^2
)}}
\nonumber \\
 &=& \sqrt{\frac{ \sigma^2_{imp} + 2T
\sigma_{imp} \frac{\partial \sigma_{imp}}{\partial T} + 2(r - q) K
T \sigma_{imp} \frac{\partial \sigma_{imp}}{\partial K} }{(1+ K
d_+ \sqrt{T} \frac{\partial \sigma_{imp}}{\partial K})^2 + K^2 T
\sigma_{imp} ( \frac{\partial^2 \sigma_{imp}}{\partial K^2}-d_+
(\frac{\partial \sigma_{imp}}{\partial K})^2 \sqrt{T})}},
\label{det_implied_locvol C}
\end{eqnarray}

\noindent with

\begin{eqnarray}
d_{\pm} = \frac{  log \frac{S(0)}{K} + (r - q \pm
\frac{\sigma^2_{imp}}{2} ) T}{\sigma_{imp} \sqrt{T}}, \nonumber
\end{eqnarray}

\begin{eqnarray}
\mathcal{N}(x) = \int_{-\infty}^{x} \frac{1}{\sqrt{2 \pi}}
e^{\frac{-z^2}{2}} dz, \nonumber
\end{eqnarray}

\begin{eqnarray}
Vega^{BS} = e^{-qT} S(0) \mathcal{N}'(d_+) \sqrt{T}, \nonumber
\end{eqnarray}

\begin{eqnarray}
\mathcal{N}'(x) = \frac{1}{\sqrt{2 \pi}} e^{\frac{-x^2}{2}}.
\nonumber
\end{eqnarray}

Using the same approach, the local volatility expression
(\ref{locvol C}) can be written in terms of implied volatilities
$\sigma_{imp}$,

\begin{align}
\sigma(T,K) =  \sqrt{\frac{ Vega^{BS} (\frac{\sigma_{imp}}{2T}+
 \frac{\partial \sigma_{imp}}{\partial T} - q K
\frac{\partial \sigma_{imp}}{\partial K} ) + K  r(0) e^{-r(0) T}
\mathcal{N}(d_-) + K P(0,T)\mathbf{E}^{Q_T}[ r(T) \textbf{1}_{
\{S(T)>K\}}]}{\frac{1}{2} K^2 Vega^{BS} (\frac{1}{\sigma_{imp} K^2
T}+ \frac{2 d_+}{\sigma_{imp} K \sqrt{T}} \frac{\partial
\sigma_{imp}}{\partial K} + \frac{\partial^2
\sigma_{imp}}{\partial K^2} + \frac{d_+ d_-}{\sigma_{imp}}
(\frac{\partial \sigma_{imp}}{\partial K})^2 )}}. \nonumber \\
\label{locvol_implied C}
\end{align}

\subsection{The Monte Carlo approach \label{Monte Carlo approach}}

In this section we present a Monte Carlo approach for the
calibration of the local volatility expression (\ref{locvol C}) or
(\ref{locvol_implied C}). More precisely, we use Monte Carlo
simulations to calculate an approximation of the expectation
$\mathbf{E}^{Q_T}[ r(T) \textbf{1}_{ \{S(T)>K\}} ]$. Therefore we
have to simulate interest rates $r(t)$ and the fund value $S(t)$
up to time $T$ starting from the actual interest rate $r(0)$ and
fund value $S(0)$ respectively. Note that in the remainder of the
paper we concentrate on the Hull and White model where
$\alpha(t)=\alpha$ and $\sigma_r(t)=\sigma_r$ are positive
constants. In \cite{HullandWhite1995b}, Hull and White remarked
that the future volatility structure implied by
(\ref{rd_HW_dynamic}) are likely to be unrealistic in the sense
that they do not conform to typical market shapes. We therefore
assume, exactly as Hull and White in \cite{HullandWhite1994A},
$\alpha(t)$ and $\sigma_r(t)$ as positive constants. In that case,
 one can exactly fit the market term structure of interest rates if
the parameter $\theta(t)$ satisfies (see \cite{Brigo-Mercurio} and
\cite{HullandWhite1994A})

\begin{eqnarray}
\theta(t) = \frac{\partial f^{mkt}(0,t)}{\partial T} + \alpha
f^{mkt}(0,t)+ \frac{\sigma_{r}^{2}}{2 \alpha^{2}} (1-e^{ -2\alpha
t}), \nonumber
\end{eqnarray}

\noindent where $f^{mkt}(0,t)$ denotes the market instantaneous
forward rate at time 0 for the maturity $t$ and where
$\frac{\partial f^{mkt}}{\partial T}$  denotes partial derivatives
of $f^{mkt}$ with respect to its second argument.\\

Since the expectation $\mathbf{E}^{Q_T}[ r(T) \textbf{1}_{
\{S(T)>K\}}]$ is expressed under the measure $Q_T$, we use the
dynamics of $S(t)$ and $r(t)$ under that measure, namely,

\begin{numcases}{}
dS(t) = (r(t) - q -  \sigma(t,S(t)) \sigma_r b(t,T) \rho_{Sr} )  S(t)  dt +  \sigma(t,S(t)) S(t)  dW^{Q_T}_{S}(t), & \label{TF dynamics1} \\
dr(t) = [\theta(t) - \alpha  r(t) - \sigma^{2}_r(t) b(t,T) ] dt +
\sigma_r dW^{Q_T}_{r}(t), & \label{TF dynamics2}
\end{numcases}

\noindent where $b(t,T)= \frac{1}{\alpha} (1-e^{-\alpha(T-t)})$.\\

Following the Monte Carlo principle, we simulate $n$ times (i.e.
$n$ scenarios) the stochastic variables $r(t)$ and $S(t)$ up to
time $T$, using $\Delta t$ as time step of discretization  and
applying an Euler scheme for example. Therefore, the expectation
is approximated by:

\begin{eqnarray}
\mathbf{E}^{Q_T}[ r(T) \textbf{1}_{ \{S(T)>K\}} ] \cong
\frac{1}{n} \sum_{i=1}^{n} r_i(T) \textbf{1}_{ \{S_i(T)>K\}},
\end{eqnarray}

\noindent where $i$ corresponds to the $i^{th}$-scenario,
$i=1,...,n$.
\\

Note that with a well-known change of variable, one can remove the
need to calculate $\theta(t)$. The idea is to rewrite the
stochastic interest rates as a sum of a stochastic and a
deterministic part (see \citep{Brigo-Mercurio}):

\begin{eqnarray}
r(t) = x(t) + \overline{x}(t),
\end{eqnarray}

\noindent where the stochastic part obeys the following dynamics:

\begin{eqnarray}
dx(t) = - (\alpha   x(t) + \sigma^{2}_r(t) b(t,T))   dt +
\sigma_{r}(t) dW^{Q_T}_{r}(t) \label{TF dynamics3}
\end{eqnarray}

\noindent and where the deterministic part obeys the dynamics :

\begin{eqnarray}
d\overline{x}(t) = (\theta(t) - \alpha   \overline{x}(t)) dt,
\end{eqnarray}

\noindent which yields to (see \citep{Brigo-Mercurio}):

\begin{eqnarray} \overline{x}(t) = f^{mkt}(0,t) +
\frac{\sigma_{r}^{2}}{2 \alpha^{2}} (1-e^{ -\alpha t})^{2}.
\nonumber
\end{eqnarray}

To illustrate the algorithm, consider the case of using an Euler
scheme for equations (\ref{TF dynamics1}) and (\ref{TF
dynamics3}),

\begin{eqnarray}
S(t_k+1) &=& S(t_{k}) + (r(t_{k}) - q - \sigma(t_{k},S(t_{k}))
\sigma_{r} b(t_{k},T) \rho_{Sr})  S(t_{k}) \Delta t +
\sigma(t_{k},S(t_{k})) S(t_{k}) \sqrt{ \Delta t} Z_{S}, \nonumber \\ 
x(t_{k+1}) &=& x(t_{k}) - [ \alpha x(t_{k}) +  \sigma^{2}_{r}
b(t_{k},T) ] \Delta t + \sigma_{r}
\sqrt{\Delta t} [ \rho_{Sr} Z_{S} + \sqrt{1-\rho_{Sr}^2} Z_{r}], \nonumber \\
r(t_{k+1}) &=& x(t_{k+1}) + \overline{x}(t_{k+1}), \label{path for
r}
\end{eqnarray}


\noindent where

\begin{eqnarray}
\overline{x}(t_k) = f^{mkt}(0,t_k) + \frac{\sigma_{r}^{2}}{2
\alpha^{2}} (1-e^{ -\alpha t_k })^{2} \ \text{and} \ b(t_k,T)=
\frac{1}{\alpha} (1-e^{-\alpha(T-t_k)}) \nonumber
\end{eqnarray}

\noindent and $Z_r$ and $Z_S$ are two independent standard normal
variables.\\

As one can see in equation (\ref{path for r}), the local
volatility function has to be known for the simulation of the path
for $S(t)$ and $r(t)$. Consequently, the only possible way to work
is forward in time. To begin, we have to determine the local
volatility function at the first time step $T=T_1$ for all strikes
$K$. At this first step we assume that the initial local
volatility is equal to the deterministic local volatility given by
equation (\ref{detreministic locvol C}). Note that this local
volatility is directly obtained by using market data (see equation
(\ref{det_implied_locvol C})). More precisely, by this choice, we
assume that for a ``small time period", interest rates are
constant and in this case, the local volatility expression
(\ref{locvol C}) reduces to (\ref{detreministic locvol C}).
Knowing that local volatility function, $S(T_1)$ and $r(T_1)$ can
be simulated. Then, the expectation $\mathbf{E}^{Q_{T_1}}[ r(T_1)
\textbf{1}_{\{S(T_1)>K\} }]$ can be computed for all $K$ by using:

\begin{eqnarray} \mathbf{E}^{Q_{T_1}}[
r(T_1) \textbf{1}_{ \{S(T_1)>K\}} ] \cong \frac{1}{n}
\sum_{i=1}^{n} r_i(T_1) \textbf{1}_{ \{S_i(T_1)>K\}}.
\end{eqnarray}

At this step we are able to build the local volatility function at
time $T_1$, $\sigma^{2}(T_1,K)$ for all strikes $K$. Following the
same procedure we can easily calibrate the local volatility at
time $T_2$ by using the local volatility obtained at time $T_1$
and also the simulated paths until time $T_1$. Following this
procedure we are able to generate the local volatility expression
up to a final date $T=T_k$.\\

\subsection{Comparison between local volatility with and without stochastic interest rates \label{section: calibration Comparison between local volatility}}

Assuming that quantities $\mathbf{Cov}^{Q_T}[r(T), \textbf{1}_{
\{S(T)>K\}}]$  are extractable from the market, it is possible to
adjust the tractable Dupire local volatility function
$\sigma_{1f}(T,K)$ coming from the one-factor Gaussian model (see
equation (\ref{detreministic locvol C})) in order to obtain the
local volatility surface which takes into account the effects of
stochastic interest rates $\sigma_{2f}(T,K)$ (i.e. equation
(\ref{locvol C})).\\

More precisely, the adjustment $\sigma_{2f}^{2}(T,K) -
\sigma_{1f}^{2}(T,K) $ is given by\footnote{Details about the
derivation of (\ref{loc vol difference}) can be received on
request. Remark that a similar derivation can be found in
\citep{Atlan} or \citep{Deelstra-Rayee} in other settings.}

\begin{eqnarray}
\sigma_{2f}^{2}(T,K) - \sigma_{1f}^{2}(T,K)  = \frac{  P(0,T)
\mathbf{Cov}^{Q_T}[r(T), \textbf{1}_{ \{S(T)>K\}} ] }{ \frac{1}{2}
K \frac{\partial^{2} C}{\partial K^{2}}},\label{loc vol
difference}
\end{eqnarray}

\noindent \noindent where $\mathbf{Cov}^{Q_T}(X,Y)$ represents the
covariance between two stochastic variables X and Y with dynamics
expressed in the $T$-forward measure $Q_T$.

\subsection{Calibrating the local volatility by mimicking stochastic volatility
models \label{section : mimicking stoch vol}}

In this subsection, we give the link between the local volatility
model and a stochastic volatility one under the assumption that
interest rates are stochastic in both models. This link gives a
way to calibrate
the local volatility from stochastic volatility models.\\

Consider the following risk neutral dynamics for the equity spot

\begin{eqnarray}
dS(t) = (r(t) - q) S(t) dt + \gamma(t,\nu(t)) S(t) dW^Q_{S}(t),
\label{DRN dynamics with stoch vol}
\end{eqnarray}

\noindent Common designs for the function $\gamma(t,\nu(t))$ are
$\nu(t)$, $ exp(\sqrt{\nu(t)})$ and $\sqrt{\nu(t)}$. The
stochastic variable $\nu(t)$  is generally modelled by  a
Cox-Ingersoll-Ross (CIR) process (as for example the Heston model
\citep{heston}) or if the function $\gamma(t,\nu(t))$ allows for
negative values for the stochastic volatility $\nu(t)$, by an
Ornstein-Uhlenbeck process (OU) (as for example the Sch\"{o}bel
and Zhu \citep{SchobelandZhu} stochastic volatility model).\\

In \citep{Atlan} and \citep{Deelstra-Rayee}, the authors have
shown that if there exists a local volatility such that the
one-dimensional probability distribution of the equity spot with
the diffusion (\ref{spot_dynamic}) is the same as the one of the
equity spot with dynamics (\ref{DRN dynamics with stoch vol}) for
every time $t$ when assuming that the risk neutral probability
measure $Q$ used in the stochastic and the local volatility
framework is the same, then the local volatility function
\mbox{$\sigma(t,S(t)=K)$} is given by the square root of the
conditional expectation under the $t$-forward measure of the
instantaneous equity stochastic spot volatility at the future time
$t$, conditional on the equity spot level $S(t)$ being equal to
$K$:

\begin{eqnarray}
\sigma(t,K) =  \sqrt{\mathbf{E}^{Q_{t}}[\gamma^2(t,\nu(t)) \mid
S(t) = K ]}. \label{resultat07}
\end{eqnarray}

\begin{Rem}
If we assume independence between the spot equity and its
volatility, the local volatility function is given by $\sigma(T,K)
= \sqrt{\mathbf{E}^{Q_T}[ \gamma^2(T,\nu(T))]}$. Depending on the
dynamics chosen for the stochastic variable $\nu(t)$, it is
sometimes possible to derive closed-form solutions for this
expectation. For example consider the following stochastic
volatility model where interest rates have Hull and White dynamics
and with Sch\"{o}bel and Zhu dynamics for the equity spot
volatility ($\nu(t)$):

\begin{numcases}{}
dS(t) = (r(t) - q)  S(t)  dt + \nu(t)  S(t)  dW^Q_{S}(t), &  \\
dr(t) = [\theta(t) - \alpha  r(t) ]  dt +
\sigma_r dW^Q_{r}(t), & \\
d \nu(t) = \kappa [\psi -\nu(t)] \ dt + \tau dW^Q_{\nu}(t).
\label{Schobel and Zhu dynamics} &
\end{numcases}

In this case, we have a closed-form solution for the local
volatility function given by

\begin{eqnarray}
\sigma(T,K) = \sqrt{\left( \nu(0) e^{- \kappa T} + (\psi -
\frac{\rho_{r \nu} \sigma_r \tau}{\alpha \kappa}) (1 - e^{- \kappa
T}) + \frac{\rho_{r \nu} \sigma_r \tau}{\alpha (\alpha+\kappa)) }
(1- e^{-(\alpha +\kappa)T}) \right)^2 + \frac{\tau^2}{2 \kappa}
(1- e^{-2 \kappa T})}. \nonumber
\end{eqnarray}

\end{Rem}


\section{Variable Annuity Guarantees \label{section Variable Annuities}}

In this section we present three different Variable Annuity
products and in section \ref{section GAO results}, we discuss the
price of all these products using local, stochastic and constant
volatility models. We first define in Subsection
\ref{section_GAO_def} the Guaranteed Annuity Option and afterwards
we define in Subsection \ref{subsection:GMIB_Rider} the Guaranteed
Minimum Income Benefit (Rider). This last product has the
particularity to be strongly dependent upon the path of the fund
value. Finally, in subsection \ref{section:Barrier GAO}, we study
two barrier type GAOs with a strong dependence upon the path of
the interest rates.

\subsection{Guaranteed Annuity Options \label{section_GAO_def}}

Consider an $x$ year policyholder who disposes at time $T$ of the
payout of his capital policy which corresponds to an amount of
money $S(T)$. A Guaranteed Annuity Option gives to the
policyholder the right to choose either an annual payment of $S(T)
g$ where $g$ is a fixed rate called the Guaranteed Annuity rate or
a cash payment equal to the equity fund value $S(T)$ at time $T$
which can be considered as an annual payment of $S(T) r_{x}(T)$,
with $r_{x}(T)$ being the market annuity payout rate defined by
$r_{x}(T) = \frac{1}{\ddot{a}_{x}(T)}$ with $\ddot{a}_{x}(T)=
\overset{\omega - (x+T)}{\underset{n=0}\sum} \phantom{.}_{n}
p_{x+T} P(T,T+n)$ where $\omega$ is the largest survival age,
$P(T,T+n)$ is the zero-coupon bond at time $T$ maturing at $T+n$
and $\phantom{.}_{n} p_{x + T}$ is the probability that the
remaining lifetime of the policyholder at time $T$ is strictly
greater than $n$. At time $T$ the value of the GAO is given by

\begin{eqnarray}
V(T) & = & max(S(T) g \ddot{a}_{x}(T) , S(T))  \label{GAO payoff 01}\\
&  = & S(T)  +  g S(T)  max(( \sum_{n=0}^{\omega - (x+T)}
\phantom{.}_{n} p_{x+T} \ P(T,T+n)) - K,0), \label{GAO_payoff}
\end{eqnarray}

\noindent where $K=\frac{1}{g}$. \\

Assuming that the mortality risk is unsystematic and independent
of the financial risk and applying the risk-neutral valuation
procedure, we can write the value of a GAO entered by an $x$-year
policyholder at time $t=0$ as

\begin{eqnarray}
V(x,0,T) & = & E^{Q}[e^{- \int_{0}^{T} r(s) ds} V(T)
1_{(\tau_{x}>T)} | \mathcal{F}_{0}] \nonumber \\
&  = & E^{Q}[e^{- \int_{0}^{T} r(s) ds} V(T)]
E^{Q}[1_{(\tau_{x}> T)}] \nonumber \\
&  = & E^{Q}[e^{- \int_{0}^{T} r(s) ds} V(T)] \phantom{.}_{T}
p_{x}, \label{GAO_equ_01}
\end{eqnarray}

\noindent where $\tau_{x}$ is a random variable which represents
the remaining lifetime of the policyholder. Substituting equation
(\ref{GAO_payoff}) in (\ref{GAO_equ_01}) leads to

\begin{eqnarray}
 V(x,0,T)  =    \phantom{.}_{T} p_{x}  E^{Q}[e^{- \int_{0}^{T} r(s)
ds}  S(T)] + C(x,0,T).
\end{eqnarray}

\noindent where

\begin{eqnarray}
C(x,0,T) = \phantom{.}_{T} p_{x} E^{Q}\left[e^{- \int_{0}^{T} r(s)
ds} g S(T) [ max((\sum_{n=0}^{\omega - (x+T)} \phantom{.}_{n}
p_{x+T} P(T,T+n)) - K,0)]\right].
\end{eqnarray}

Under the usual assumption of absence of arbitrage opportunities,
the discounted value of the risky fund $e^{- \int_{0}^{T} r(s) ds}
S(T)$ is a martingale under the risk neutral measure $Q$ and
therefore the value of the GAO becomes

\begin{eqnarray}
 V(x,0,T)  =    \phantom{.}_{T} p_{x}  S(0) + C(x,0,T). \label{equationforV}
\end{eqnarray}

The first term in equation (\ref{equationforV}) is a constant and
therefore, in the literature, one studies generally only the
second term $C(x,0,T)$. In \cite{Ballotta} and
\cite{van_Haastrecht_GAO}, the authors define $C(x,0,T)$ as the
GAO total value. In this paper we keep the same terminology. More
precisely, in section \ref{section GAO results}, when we compare
GAO total values obtained in different models, we compare values
obtained for $C(x,0,T)$.\\

To derive analytical expressions for $C(x,0,T)$ in the
Black-Scholes Hull-White and the Sch\"{o}bel-Zhu Hull-White models
it is more convenient to work under the measure $Q_S$ (where the
numeraire is the fund value $S$), rather than under the risque
neutral measure $Q$ (see \cite{Ballotta} and
\cite{van_Haastrecht_GAO}). By the density process $\xi_{T} =
\frac{dQ_{S}}{dQ}|_{\mathcal{F}_{T}} = e^{- \int_{0}^{T} r(s) ds}
\frac{S(T)}{S(0)}$ a new probability measure $Q_S$ equivalent to
the measure $Q$ is defined, see e.g.\ Geman et al. \cite{Geman}.
Under this new measure $Q_S$, the $C(x,0,T)$ value becomes

\begin{eqnarray}
 C(x,0,T) = \phantom{.}_{T} p_{x} g S(0) E^{Q_S}[(\sum_{n=0}^{\omega - (x+T)} \phantom{.}_{n} p_{x+T}  P(T,T+n) -
K)^{+}]. \label{GAO_under_QS2}
\end{eqnarray}

Under $Q_S$ one has the following model dynamics

\begin{eqnarray}
dS(t) &=& [r(t) - q + \sigma^{2}(t,S(t))] S(t) dt + \sigma(t,S(t)) S(t) dW^{Q_S}_{S}(t), \\
dx(t) &=& [- \alpha  x(t) + \rho_{rS} \sigma_r \sigma(t,S(t))] dt
+ \sigma_{r} dW^{Q_S}_{r}(t). \label{x_dyn_underQS}
\end{eqnarray}

The zero-coupon bond $P(T,T+n)$ in the Gaussian Hull and White
one-factor model with $\alpha$ and $\sigma_r$ constant has the
following expression (see e.g. \citep{Brigo-Mercurio})

\begin{eqnarray}
P(T,T+n) = A(T,T+n)e^{-b(T,T+n)x(T)}, \label{zero-coupon}
\end{eqnarray}

\noindent where

\begin{eqnarray}
 A(T,T+n) & = &  \frac{P^{mkt}(0,T+n)}{P^{mkt}(0,T)} e^{-\frac{1}{2}[V(0,T+n)-V(0,T)-V(T,T+n)]},\nonumber \\
b(T,T+n) &=& \frac{1}{\alpha} (1 - e^{-\alpha(n)}), \nonumber \\
V(t_1,t_2) &=& \frac{\sigma^2_{r}}{\alpha^2} [t_2-t_1+
\frac{2}{\alpha} e^{-\alpha(t_2-t_1)}-\frac{1}{2\alpha}
e^{-2\alpha(t_2-t_1)}-\frac{3}{2\alpha}]. \nonumber
\end{eqnarray}

Substituting the expression (\ref{zero-coupon}) in equation
(\ref{GAO_under_QS2}) leads to the following pricing expression
for $C(x,0,T)$ under $Q_S$

\begin{eqnarray}
 C(x,0,T) = \phantom{.}_{T} p_{x} g S(0)  E^{Q_S}[(
\sum_{n=0}^{\omega - (x+T)} \phantom{.}_{n} p_{x+T}
A(T,T+n)e^{-b(T,T+n)x(T)} - K)^{+}]. \label{GAO_under_QS}
\end{eqnarray}

Note that when pricing GAO using a local volatility model, one has
to use some numerical methods like Monte Carlo simulations. The
calculation of the price can be based on the equation
(\ref{GAO_under_QS}) but it is also convenient to work under the
$T$-forward measure $Q_T$, where the dynamics of $S(t)$ and $r(t)$
are given by (\ref{TF dynamics1}) and (\ref{TF dynamics2})
respectively and where the GAO value is given by

\begin{eqnarray}
 C(x,0,T) = \phantom{.}_{T} p_{x} g P(0,T)  E^{Q_T}[S(T)(
\sum_{n=0}^{\omega - (x+T)} \phantom{.}_{n} p_{x+T}
A(T,T+n)e^{-b(T,T+n)x(T)} - K)^{+}]. \label{GAO_under_QT}
\end{eqnarray}

Since in both cases one has to compute the local volatility value
at each time step, both methods are equivalent in time machine
consumption.


\subsection{Guaranteed Minimum Income Benefit (Rider) \label{subsection:GMIB_Rider}}

Guaranteed Minimum Income Benefit (GMIB) is the term used in North
America for an analogous product of a GAO in Europe (see
\cite{Hardy}). GAO and GMIB payoffs are usually slightly different
but they have in common that they are both maturity guarantees in
the form of a guaranteed minimum income on
the annuitization of the maturity payout.\\



There exist many different guarantee designs for GMIB. The
policyholders can for example choose between a life annuity or a
fixed duration annuity; or choose an annual growth rate guarantee
for the fund, or a Withdraw option, etc. (for more details see
\cite{Hardy} and \cite{GMIB}). In this paper we focus on the
valuation of an ``exotic GMIB", namely a GMIB Rider. \\

A GMIB Rider (based on examples given in \cite{GMIB} and in
\cite{Hardy2}) gives the $x$ year policyholder the right to choose
at the date of annuitization $T$ between 3 guarantees: an annual
payment of $g S(0) (1+ r_g)^T $ where $r_g$ is a guaranteed annual
rate; an annual payment of $g \ \underset{n}{max}(S(n)) $,
$n=1,2,...,T$ where $S(n)$, $n=1,2,...,T$ are the anniversary
values of the fund or a cash payment equal to the equity fund
value $S(T)$ at maturity $T$. Therefore, at time $T$ the value of
the GMIB Rider is given by

\begin{eqnarray}
V(T) = max(  S(0) (1+ r_g)^T g \ddot{a}_{x}(T) , \underset{n \in
A}{max}(S(n)) g \ddot{a}_{x}(T) , S(T)), \label{GMIB RIDER payoff}
\end{eqnarray}

\noindent where $A$ is the set of anniversary dates $A= \{
1,2,...,T \}$. The valuation of this product in the Black-Scholes
Hull-White model has been studied by e.g.\ Marshall et al.\ in \cite{Hardy2}.\\

Assuming that the mortality risk is unsystematic and independent
of the financial risk, we can write the value of a GMIB Rider
entered by an $x$-year policyholder at time $t=0$ as

\begin{eqnarray}
V(x,0,T) & = &  E^{Q}[e^{- \int_{0}^{T} r(s) ds} V(T)]
\phantom{.}_{T}
p_{x} \nonumber \\
&  = & P(0,T) E^{Q_T}[V(T)] \phantom{.}_{T} p_{x}. \label{GMIB
RIDER_equ_01}
\end{eqnarray}

The GMIB Rider payoff is path-dependent and more complicated than
a pure GAO. There is no closed-form expression in the BSHW neither
in the SZHW model nor in the LVHW model. Consequently one has to
use numerical methods to evaluate the expectation in equation
(\ref{GMIB RIDER_equ_01}) in all models we consider. In Subsection
\ref{section:GMIB_results}, we compare GMIB Rider values given by
using the LVHW model, the SZHW model and the constant volatility
BSHW model using a Monte Carlo approach.

\subsection{Barrier GAOs \label{section:Barrier GAO}}

In this section, we introduce path-dependent GAOs which are
barrier type options. The first one is a ``down-and-in GAO" which
becomes activated only if the interest rates reach a downside
barrier level $B$. This product is interesting for buyers since it
protects against low interest rates at the retirement age and has
a smaller price than the pure GAO price. The second exotic GAO we
study here is the ``down-and-out GAO" which becomes deactivated if
the interest rates reach a downside barrier level $B$. This
barrier option gives protection to insurers against low market of
interest rates. For example it allows to avoid situations as it
occurs in the UK where insurers have sold products with guarantee
rate $g$ of 11\% while the market rates are now really
low,\footnote{In United Kingdom in the 1970's and 1980's the most
popular Guaranteed Annuity rate proposed by UK life insurers was
about 11\%  (see Bolton et al. \cite{Bolton}).} resulting in
considerable losses for insurance companies. To our knowledge, no
closed-form solutions exist for the price of these path-dependent
GAOs in the BSHW model nor in the SZHW and LVHW models. Therefore
we
will also price these path-dependent derivatives by using a Monte Carlo simulation approach.\\

Under the measure $Q_S$, the price of the ``down-and-out GAO" is
given by the following expression

\begin{eqnarray}
 C^{DO}(x,0,T) = \phantom{.}_{T} p_{x} g S(0)  E^{Q_S}[(
\sum_{n=0}^{\omega - (x+T)} \phantom{.}_{n} p_{x+T}
A(T,T+n)e^{-b(T,T+n)x(T)} - K)^{+} \mathrm{1}_{ (x(t)>B, \ 0< t
\leq T)}]. \label{DO_GAO_equation}
\end{eqnarray}

\begin{figure}
\begin{center}
\includegraphics[height=7cm]{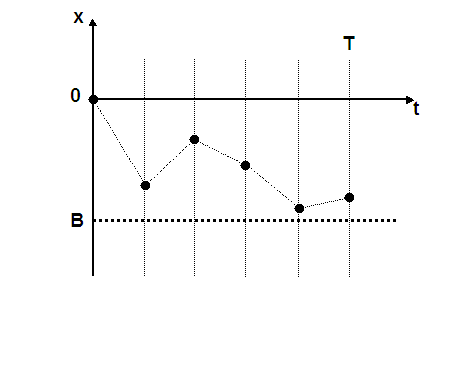}
\caption{Example of path realization}\label{fig:psurv}
\end{center}
\end{figure}

When pricing barrier options with Monte Carlo simulations it is
possible to miss some barrier hitting realization between two time
steps. Consider the particular path realization of Figure
\ref{fig:psurv}. There are five time steps and none of the five
underlying realizations have breached the barrier and therefore
the path is going to count in the payoff sum. Following the idea
presented in \cite{Shevchenko}, we should weight the actualized
payoff by a certain factor accounting for the probability of
breaching the barrier between the discrete time points. As a
weighting factor we can use the product over all time intervals
$\Delta_t^i$ of the survival probability of the option,

\begin{eqnarray}
 w = \prod_{i=1}^{n} P_{surv}(\Delta_t^i).
\end{eqnarray}

Assuming that the volatility of the equity is constant over the
interval $\Delta_t^i$, it is possible to derive an analytical
expression for $P_{surv}(\Delta_t^i)$ applicable in all different
models treated in this paper. More precisely, under this
assumption, $x(t+\Delta_t^i)$ is normally distributed with mean
$\mu_x$ and variance  $\sigma^2_x$ given by

\begin{eqnarray}
 \mu_x &=& \frac{\rho_{Sr} \sigma_r \sigma_S}{\alpha} (1-e^{-\alpha \Delta_t^i}),\nonumber \\
\sigma^2_x &=& \frac{ \sigma^2_r }{2\alpha} (1-e^{-2 \alpha
\Delta_t^i}).
\end{eqnarray}

Using the well-known reflection principle (see \cite{Shreve}), we
have the analytical expression for $P_{surv}(\Delta_t^i)$

\begin{eqnarray}
 P_{surv}(\Delta_t^i) &=& P(x(s) > B, \ t < s
\leq t + \Delta_t^i) \\
&=& \mathcal{N}\left(\frac{-B+\mu_x \Delta_t^i}{\sigma_x
\sqrt{\Delta_t^i}}\right)-e^{(\frac{-2B\mu_x }{\sigma^2_x})}
\mathcal{N}\left(\frac{B+\mu_x \Delta_t^i}{\sigma_x
\sqrt{\Delta_t^i}}\right).
\end{eqnarray}

The last assumption about the volatility of the equity is in
contradiction with the nature of the LVHW and the SZHW models.
However, as time intervals become smaller, this assumption is more
and more justified.\\

The price of the ``down-and-in GAO" can easily be computed from
the price of the ``down-and-out GAO" and the ``pure GAO" by using
the following relation

\begin{eqnarray}
 C^{DI}(x,0,T) = C(x,0,T) - C^{DO}(x,0,T).  \label{DI_GAO_equation}
\end{eqnarray}

Note that for knock-in type options, the survival probability is
readily derived from the one coming from the corresponding
knock-out type option by complementarity ($P_{surv}^{KI} =
1-P_{surv}^{KO}$).\\

\section{Numerical results \label{section_numerical results}}

In this section we study the contribution of using a local
volatility model with Hull and White stochastic interest rates
(LVHW) (introduced in Section \ref{section:the model}) to the
pricing of Variable Annuity Guarantees. More precisely, we compare
GAO, GMIB Rider and two barrier types GAO prices obtained by using
the LVHW model to those obtained with the Sch\"{o}bel-Zhu
Hull-White (SZHW) stochastic volatility model and the
Black-Scholes with Hull and White stochastic interest rates
(BSHW). For a fair analysis, one first has to calibrate these
three models to the same options market data. For this end, we
have used the same data as in \citep{van_Haastrecht_GAO}. More
precisely, the equity components (fund) of the Variable Annuity
Guarantees considered are on one hand the EuroStoxx50 index (EU)
and on the other hand the S\&P500 index (US). In the following
subsection we explain in details the calibration of the three
models. We summarize the calibration of the interest rate
parameters and the calibration of the SZHW and the BSHW made in
\citep{van_Haastrecht_GAO} and then we explain the calibration of
the local volatility surface for the equity component in our LVHW
model. In Subsection \ref{section GAO results} we compare GAO
values obtained by using the LVHW model with the SZHW and the BSHW
prices studied in \citep{van_Haastrecht_GAO}. In Subsection
\ref{section:GMIB_results} and Subsection \ref{section: Barrier
GAO results}, we do the same study for path-dependent Variable
Annuity Guarantees namely GMIB Riders and barrier type GAOs.

\subsection{Calibration to the Vanilla option's Market \label{section Calibration
results}}

In order to compare LVHW GAO results to the ones obtained in the
BSHW and the SZHW models presented in van Haastrecht et al.
\citep{van_Haastrecht_GAO} we are using Hull and White parameters
and the implied volatility curve they have used\footnote{We would
like to thank A. van Haastrecht, R. Plat and A. Pelsser for
providing us the Hull and White parameters and interest rate curve
data they used in \cite{van_Haastrecht_GAO}.}. In
\citep{van_Haastrecht_GAO}, interest rate parameters are
calibrated to EU and US swaption markets using swaption mid prices
of the 31st of July 2007. Moreover, the effective 10 years
correlation between the log equity returns and the interest rates
is determined by time series analysis of the 10-year swap rate and
the log returns of the EuroStoxx50 index (EU) and the S\&P500
index (US) over the period from February 2002 to July 2007 and
turns out to be resp. $34.65 \%$ and $14.64 \%$ for the EU  and
the US market. The equity parameters in the SZHW and BSHW models
are calibrated by using vanilla option prices on the EuroStoxx50
and S\&P500 index obtained from the implied volatility service of
MarkIT\footnote{A financial data provider, which provides (mid)
implied volatility quotes by averaging quotes from
a large number of issuers.}.\\

In \citep{van_Haastrecht_GAO} the authors have calibrated the
equity model to market option prices maturing in 10 years time. In
the LVHW model, the equity volatility is a local volatility
surface and the calibration consists in building this surface
using equation (\ref{locvol_implied C}). This calibration
procedure uses the whole implied volatility surface and returns a
local volatility for all strikes and all maturities. In this paper
we consider three different cases. In a first case, we assume that
the implied volatility is constant with respect to the maturity
($\frac{\partial \sigma_{imp}}{\partial T}=0$) and in that case we
denote the local volatility model by LVHW1. The second case
denoted by LVHW2 considers an increasing term structure
($\frac{\partial \sigma_{imp}}{\partial T}=0.01$) and finally we
consider the case LVHW3 where the implied volatility is a
decreasing function of the maturity ($\frac{\partial
\sigma_{imp}}{\partial T}=-0.003$). Note that for the aid of a
fair comparison between the models, we have always kept the same
volatility smile at time $T=10$. A plot of the three complete
implied volatility surfaces and the resulting market call option
prices can be found in Appendix \ref{appendix_graphics} (see
Figures \ref{fig:US_impvoll_dsigdtnul},
\ref{fig:US_mkt_call_dsigdtnul},
\ref{fig:US_impvoll_dsigdtpositif},
\ref{fig:US_mkt_call_dsigdtpositif},
\ref{fig:US_impvoll_dsigdtnegatif},
\ref{fig:US_mkt_call_dsigdtnegatif} for US data and
\ref{fig:EUR_impvoll_dsigdtnul}, \ref{fig:EUR_mkt_call_dsigdtnul},
\ref{fig:EUR_impvoll_dsigdtpositif},
\ref{fig:EUR_mkt_call_dsigdtpositif},
\ref{fig:EUR_impvoll_dsigdtnegatif},
\ref{fig:EUR_mkt_call_dsigdtnegatif} for EUR data). Following the
Monte Carlo approach given in Section \ref{Monte Carlo approach},
we have found the corresponding local volatility surface, see
namely Figure \ref{fig:US_locvol_fig_stoch_dsigdtnul},
\ref{fig:US_locvol_fig_stoch_dsigdtpositif},
\ref{fig:US_locvol_fig_stoch_dsigdtnegatif} for the US data and
\ref{fig:EUR_locvol_fig_stoch_dsigdtnul},
\ref{fig:EUR_locvol_fig_stoch_dsigdtpositif},
\ref{fig:EUR_locvol_fig_stoch_dsigdtnegatif} for the EUR data.\\

\begin{table}\footnotesize
\begin{center}
{\begin{tabular}{|c|c|c|c|c|} \hline
\multicolumn{ 5}{|c|}{{\bf Implied volatility, 10-year call options, US}} \\
\hline
{\bf strike} & {\bf Market} & {\bf BSHW} & {\bf SZHW} & {\bf LVHW1 ($\pm$ 95\% interval)} \\
\hline
        80 &    27.50\% &    25.80\% &    27.50\% &    27.503\% ($\pm$ 0.01778 \%)\\
\hline
        90 &    26.60\% &    25.80\% &    26.60\% &    26.601\% ($\pm$ 0.01745 \%) \\
\hline
        95 &    26.20\% &    25.80\% &    26.20\% &    26.198\% ($\pm$ 0.01693 \%) \\
\hline
       100 &    25.80\% &    25.80\% &    25.80\% &    25.800\% ($\pm$ 0.01631 \%) \\
\hline
       105 &    25.40\% &    25.80\% &    25.40\% &    25.397\% ($\pm$ 0.01501 \%) \\
\hline
       110 &    25.00\% &    25.80\% &    25.00\% &    24.998\% ($\pm$ 0.01432 \%) \\
\hline
       120 &    24.30\% &    25.80\% &    24.40\% &    24.333\% ($\pm$ 0.01325 \%) \\
\hline
\multicolumn{ 5}{|c|}{{\bf Implied volatility, 10-year call options, EUR}} \\
\hline
{\bf strike} & {\bf Market} & {\bf BSHW} & {\bf SZHW} & {\bf LVHW1 ($\pm$ 95\% interval)} \\
\hline
        80 &    27.80\% &    26.40\% &    27.90\% &    27.826\% ($\pm$ 0.0214 \%) \\
\hline
        90 &    27.10\% &    26.40\% &    27.10\% &    27.103\% ($\pm$ 0.0200 \%) \\
\hline
        95 &    26.70\% &    26.40\% &    26.70\% &    26.699\% ($\pm$ 0.0194 \%) \\
\hline
       100 &    26.40\% &    26.40\% &    26.40\% &    26.396\% ($\pm$ 0.0189 \%) \\
\hline
       105 &    26.00\% &    26.40\% &    26.00\% &    25.999\% ($\pm$ 0.0181 \%) \\
\hline
       110 &    25.70\% &    26.40\% &    25.70\% &    25.702\% ($\pm$ 0.0173 \%) \\
\hline
       120 &    25.10\% &    26.40\% &    25.10\% &    25.101\% ($\pm$ 0.0165 \%) \\
\hline
\end{tabular}
\caption{Comparison of the implied volatility curve (for 10-year
call options) generated by the SZHW, the BSHW and the LVHW1 models
after being calibrated over the market implied volatility. US and
European market implied volatilities as well as the BSHW and SZHW
results are coming from \cite{van_Haastrecht_GAO}.
\label{tab:calibration_results}}}
\end{center}
\end{table}

\begin{figure}
\begin{center}
\includegraphics[height=5 cm]{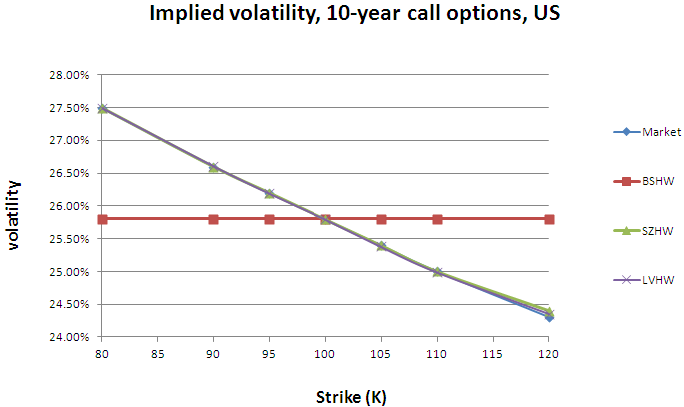}
\includegraphics[height=5 cm]{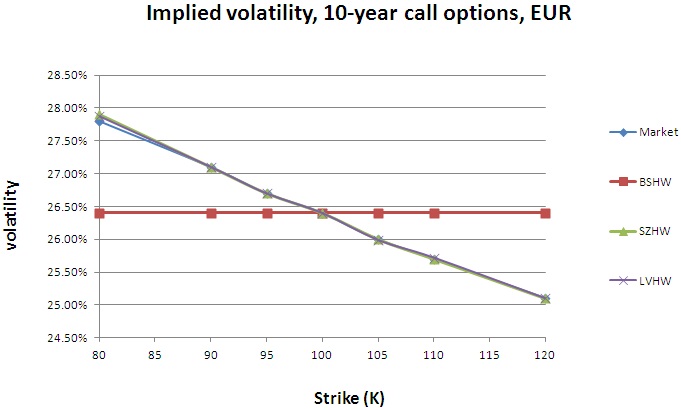}
\caption{Comparison of the market implied volatility curve (for
10-year call options) and the three implied volatility curves
coming from the SZHW, the BSHW and the LVHW1 models after
calibration.}\label{fig:models_smile}
\end{center}
\end{figure}

Stochastic volatility and local volatility models are both able to
reproduce the market smile. For example, in Table
\ref{tab:calibration_results}, we compare the market implied
volatility (for a range of seven different strikes and a fixed
maturity $T=10$) with the calibrated volatility of each of the
three models. The volatility curves of Table
\ref{tab:calibration_results} generated by the market and these
three models in the US and European 10 years maturity vanilla
options market are presented in Figure \ref{fig:models_smile}. We
notice that the local volatility model (LVHW) and the stochastic
volatility model (SZHW) are both well calibrated since they are
able to generate the Smile/Skew quite close to the market one.
Note that, the LVHW1 implied volatilities are extracted from the
European call values obtained by Monte Carlo simulations.
Therefore, in the last column of Table
\ref{tab:calibration_results}, one also gives the corresponding
$\pm$ 95\% confidence interval. However, contrarily to the SZHW
model, the LVHW model has to be calibrated over the whole implied
volatility surface and as we will see in Subsection \ref{section
GAO results}, \ref{section:GMIB_results} and \ref{section: Barrier
GAO results}, this fact has an impact to the price of Variable
Annuity Guarantees.\\

\subsection{A numerical comparison between local volatility with and without stochastic interest rates \label{section: numerical results Comparison between local volatility}}

In Section \ref{section: calibration Comparison between local
volatility}, equation (\ref{loc vol difference}) gives the
difference between the tractable Dupire local volatility function
$\sigma_{1f}(T,K)$ coming from the one-factor Gaussian model and
the one which takes into account the effects of stochastic
interest rates $\sigma_{2f}(T,K)$.\\

Note that the difference between $\sigma_{2f}(T,K)$ and
$\sigma_{1f}(T,K)$ can be derived in terms of implied volatilities
($\sigma_{imp}$) by using equation (\ref{locvol_implied C}) and
(\ref{det_implied_locvol C}) and leads to

\begin{eqnarray}
\sigma_{2f}^{2}(T,K) -  \sigma_{1f}^{2}(T,K)=  \frac{
 K  r(0) e^{-r(0) T}
\mathcal{N}(d_-) - Vega^{BS} r(0) K \frac{\partial
\sigma_{imp}}{\partial K} + K P(0,T)\mathbf{E}^{Q_T}[ r(T)
\textbf{1}_{ \{S(T)>K\}}]}{\frac{1}{2} K^2 Vega^{BS}
(\frac{1}{\sigma_{imp} K^2 T}+ \frac{2 d_+}{\sigma_{imp} K
\sqrt{T}} \frac{\partial \sigma_{imp}}{\partial K} +
\frac{\partial^2 \sigma_{imp}}{\partial K^2} + \frac{d_+
d_-}{\sigma_{imp}} (\frac{\partial \sigma_{imp}}{\partial K})^2 )}
\end{eqnarray}

In Figures \ref{fig:US_locvol_fig_stoch_dsigdtnul},
\ref{fig:US_locvol_fig_stoch_dsigdtpositif},
\ref{fig:US_locvol_fig_stoch_dsigdtnegatif},
\ref{fig:US_locvol_fig_cst_dsigdtnul},
\ref{fig:US_locvol_fig_cst_dsigdtpositif},
\ref{fig:US_locvol_fig_cst_dsigdtnegatif},
\ref{fig:US_locvol_fig_diff_dsigdtnul},
\ref{fig:US_locvol_fig_diff_dsigdtpositif} and
\ref{fig:US_locvol_fig_diff_dsigdtnegatif}, we have plotted the
local volatility coming from the stochastic interest rate
framework ($\sigma_{2f}(T,K)$), the tractable Dupire local
volatility ($\sigma_{1f}(T,K)$) and the surface generated by the
difference between these two local volatilities in the US case and
in Figures \ref{fig:EUR_locvol_fig_stoch_dsigdtnul},
\ref{fig:EUR_locvol_fig_stoch_dsigdtpositif},
\ref{fig:EUR_locvol_fig_stoch_dsigdtnegatif},
\ref{fig:EUR_locvol_fig_cst_dsigdtnul},
\ref{fig:EUR_locvol_fig_cst_dsigdtpositif},
\ref{fig:EUR_locvol_fig_cst_dsigdtnegatif},
\ref{fig:EUR_locvol_fig_diff_dsigdtnul},
\ref{fig:EUR_locvol_fig_diff_dsigdtpositif} and
\ref{fig:EUR_locvol_fig_diff_dsigdtnegatif}, the analogues in the EUR case.\\


\subsection{GAO results \label{section GAO
results}}

In this section, we study the impact generated to GAO values by
using the LVHW model with respect to prices given by using the
BSHW and the SZHW models computed in \citep{van_Haastrecht_GAO}.
We make therefore the same assumptions as in
\citep{van_Haastrecht_GAO}, namely that the policyholder is 55
years old and that  the retirement age is 65 (i.e. the maturity
$T$ of the GAO option is 10 years). The fund value at time $0$,
$S(0)$ is assumed to be 100. The survival rates are based on the
PNMA00\footnote{Available at
http://www.actuaries.org.uk/research-and-resources/pages/00-series-mortality-tables-assured-lives-annuitants-and-pensioners}
table of the Continuous Mortality Investigation (CMI) for male
pensioners.\\

\begin{figure}[ht]
\begin{center}
\includegraphics[height=4.5cm]{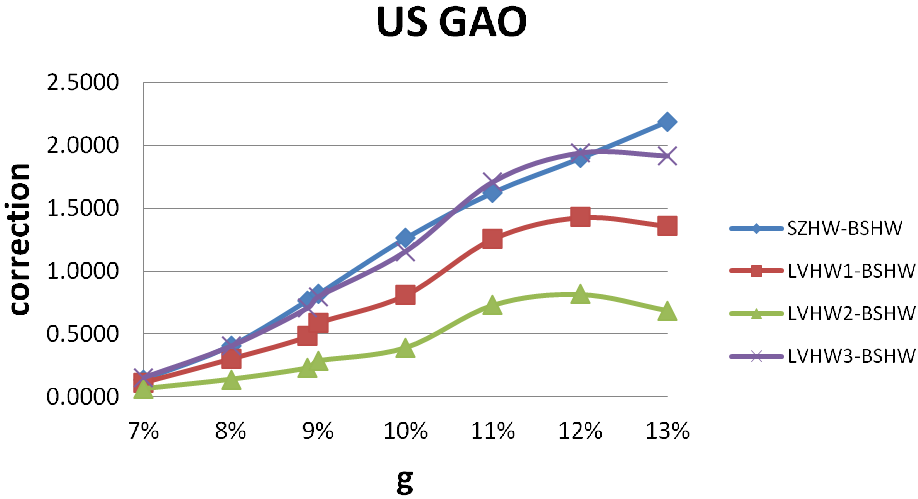}
\includegraphics[height=4.5cm]{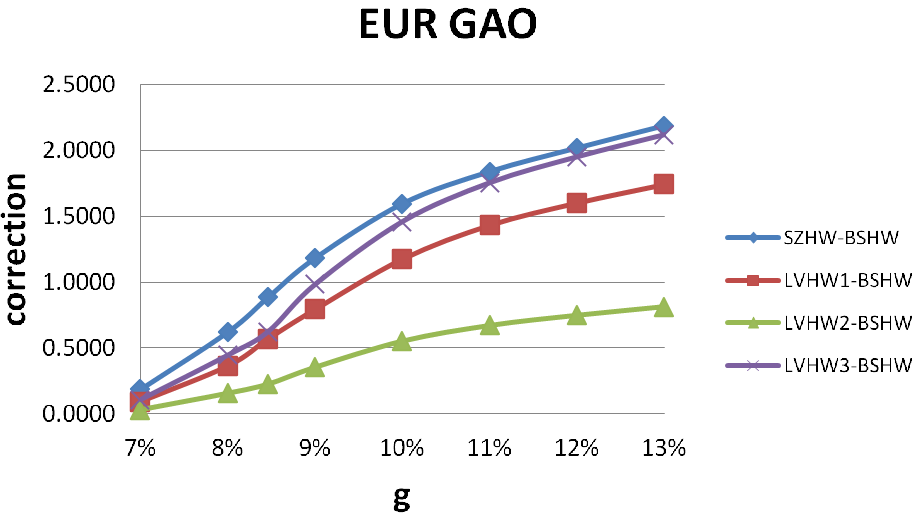}
\caption{Graphical representation of the LVHW1, LVHW2, LVHW3 and
the SZHW model corrections with respect to the BSHW model for
different guaranteed annuity rates $g$.}\label{fig:corrections}
\end{center}
\end{figure}

In Table \ref{tab:GAO_results} we give prices obtained for the GAO
using the LVHW model case 1 (LVHW1), the LVHW model case 2
(LVHW2), the LVHW model case 3 (LVHW3),  the SZHW and the BSHW
models for different guaranteed rates $g$. The results for the
SZHW and BSHW models are obtained using the closed-form expression
derived in resp. \cite{van_Haastrecht_GAO} and \cite{Ballotta},
which can be found in Appendix \ref{Appendix_closed_form} for the
convenience of the reader. Note that GAO prices presented in this
section given by the BSHW and SZHW models are slightly different
than those presented in \cite{van_Haastrecht_GAO}. These
differences come from the interpolation method we use for
constructing the zero coupon bond curve. As it was pointed out in
\cite{van_Haastrecht_GAO}, GAO prices are sensitive to the
interest rate curve and a small change in the zero coupon bond
curve induces changes in GAO prices. The results for the LVHW
model are obtained by using Monte Carlo simulations (100 000
simulations and 5000 steps). Table \ref{tab:corrections} and
Figure \ref{fig:corrections} show the corrections induced by the
LVHW and the SZHW models with respect to the BSHW
model\footnote{More precisely, the LVHW correction is the
difference between the LVHW price and the BSHW price.
Similarly, the SZHW correction is the difference between the SZHW price and the BSHW price.}.\\

\begin{table}\footnotesize
\begin{center}
{
\begin{tabular}{|c|c|c|c|c|c|c|c|c|}
\hline
                              \multicolumn{ 9}{|c|}{{\bf GAO Total Value US}} \\
\hline
  {$g$} & {\bf BSHW } & {\bf SZHW} & {\bf LVHW1} & {\bf SE} & {\bf LVHW2} & {\bf SE} & {\bf LVHW3 } & {\bf SE} \\
\hline
       7\% &   \phantom{0}0.906860 &   \phantom{0}1.045977 &   \phantom{0}1.021760  &  0.013730 &   \phantom{0}0.970480 & 0.013374 &   \phantom{0}1.059320  &  0.014154 \\
\hline
       8\% &   \phantom{0}3.160037 &   \phantom{0}3.567810  &  \phantom{0}3.463430 & 0.026539 & \phantom{0}3.299570 & 0.025895 &   \phantom{0}3.562870  &  0.027142  \\
\hline
      8.88\% &   \phantom{0}7.101917 & \phantom{0}7.869198 & \phantom{0}7.584480 & 0.039751 & \phantom{0}7.332270 &  0.038987 & \phantom{0}7.808470  &   0.040443  \\
\hline
       9\% &      \phantom{0}7.738402 & \phantom{0}8.555375 & \phantom{0}8.326130 & 0.041560 & \phantom{0}8.024770 &   0.040791 & \phantom{0}8.533410 & 0.042259  \\
\hline
      10\% &      14.880173 & 16.141393 & 15.690000  & 0.055644 & 15.271200 & 0.054907 & 16.034900 & 0.056342  \\
\hline
      11\% &      23.643769 &  25.267434 & 24.900900 & 0.066981 & 24.374900 & 0.066285 & 25.350600  & 0.067621 \\
\hline
      12\% &      33.689606 & 35.586279 & 35.117300 & 0.075737 & 34.507000  & 0.075030 & 35.626800 & 0.076333  \\
\hline
      13\% &      44.382228 & 46.570479 & 45.742500  & 0.082976 & 45.070200 & 0.082227 & 46.296100 & 0.083563 \\
\hline
                                \multicolumn{ 9}{|c|}{{\bf GAO Total Value EUR}} \\
\hline
  {$g$} & {\bf BSHW } & {\bf SZHW} & {\bf LVHW1} & {\bf SE} & {\bf LVHW2} & {\bf SE} & {\bf LVHW3 } & {\bf SE} \\
\hline
       7\% &   \phantom{0}0.395613 &    \phantom{0}0.583057 &   \phantom{0}0.486412  &  0.007961  &  \phantom{0}0.428838 & 0.007326 & \phantom{0}0.502376   & 0.008006 \\
\hline
       8\% &   \phantom{0}2.259404  &  \phantom{0}2.882720 &   \phantom{0}2.622240 & 0.019585 & \phantom{0}2.417440  &  0.018578   & \phantom{0}2.704880 &   0.019687  \\
\hline
      8.46\% &  \phantom{0}4.095928 & \phantom{0}4.981611 & \phantom{0}4.664650 &    0.026196  &  \phantom{0}4.323210  &  0.025098 & \phantom{0}4.719230 &   0.026313  \\
\hline
       9\% &      \phantom{0}7.210501  &  \phantom{0}8.393904  &  \phantom{0}8.004080 &   0.033921 & \phantom{0}7.565690 &   0.032806 & \phantom{0}8.193780 &   0.034034 \\
\hline
      10\% &      15.421258 &  17.017295 &  16.597400 &  0.045658 & 15.974700 &  0.044657  &  16.880900 & 0.045707   \\
\hline
      11\% &      25.530806 &  27.369899 & 26.964500 &  0.053381  &  26.202900 & 0.052448  &  27.285900 &  0.053395 \\
\hline
      12\% &      36.282533 &  38.300598  & 37.885600 & 0.058946 &   37.031900  & 0.057988 &  38.234200  & 0.058954 \\
\hline
      13\% &     47.164918 & 49.351551  & 48.908600 &  0.063960 & 47.979600 &  0.062924 & 49.284400 &  0.063964 \\
\hline
\end{tabular}
\caption{Comparison of GAO total values in the BSHW, the SZHW and
the LVHW1, LVHW2 and LVHW3 models for different guaranteed annuity
rates $g$. The rates $g$ of $8.88 \%$ and $8.46 \%$ correspond to
the at-the-money guaranteed annuity rates in the US and EUR market
respectively. \label{tab:GAO_results}}}
\end{center}
\end{table}

\begin{table}\footnotesize
\begin{center}
{\begin{tabular}{|c|c|c|c|c|} \hline
                    \multicolumn{ 5}{|c|}{{\bf US Correction}} \\
\hline
  {$g$} & {\bf SZHW-BSHW} & {\bf LVHW1-BSHW} & {\bf LVHW2-BSHW} & {\bf LVHW3-BSHW} \\
\hline
       7\% &       0.1391&  0.1149&  0.0636&  0.1525 \\
\hline
       8\% &       0.4078&  0.3034&  0.1395&  0.4028 \\
\hline
    8.88\% &       0.7673&  0.4826&  0.2304&  0.7066 \\
\hline
       9\% &       0.8170&  0.5877&  0.2864&  0.7950 \\
\hline
      10\% &       1.2612&  0.8098&  0.3910&  1.1547 \\
\hline
      11\% &      1.6237& 1.2571&  0.7311&  1.7068 \\
\hline
      12\% &       1.8967&  1.4277&  0.8174&  1.9372  \\
\hline
      13\% &       2.1883&  1.3603&  0.6880&  1.9139 \\
\hline
                   \multicolumn{ 5}{|c|}{{\bf EUR Correction}} \\
\hline
  {$g$} & {\bf SZHW-BSHW} & {\bf LVHW1-BSHW} & {\bf LVHW2-BSHW} & {\bf LVHW3-BSHW} \\
\hline
       7\% &        0.1874&  0.0908&  0.0332&  0.1068 \\
\hline
       8\% &        0.6233&  0.3628&  0.1580 & 0.4455 \\
\hline
    8.46\% &       0.8857&  0.5687&  0.2273 & 0.6233 \\
\hline
       9\% &       1.1834&  0.7936&  0.3552 & 0.9833 \\
\hline
      10\% &       1.5960 & 1.1761&  0.5534&  1.4596 \\
\hline
      11\% &       1.8391&  1.4337&  0.6721&  1.7551 \\
\hline
      12\% &       2.0181 & 1.6031 & 0.7494 & 1.9517 \\
\hline
      13\% &       2.1866&  1.7437&  0.8147&  2.1195 \\
\hline
\end{tabular}
\caption{Corrections generated by the LVHW1, LVHW2, LVHW3 and the
SZHW models with respect to the BSHW model for different
guaranteed annuity rates $g$. \label{tab:corrections}}}
\end{center}
\end{table}

\begin{figure}[ht]
\begin{center}
\includegraphics[height=4.5cm]{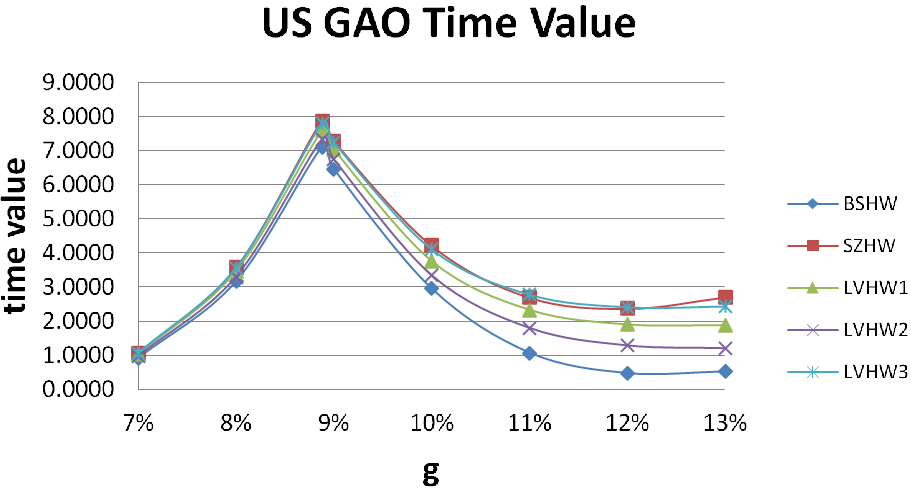}
\includegraphics[height=4.5cm]{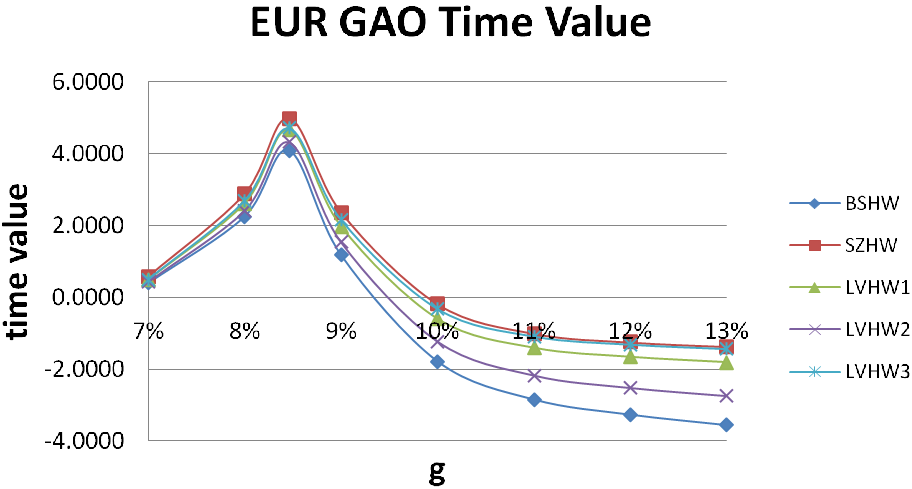}
\caption{Graphical representation of GAO time values of the BSHW,
the SZHW and the LVHW1, LVHW2 and LVHW3 models for different
guaranteed annuity rates $g$.}\label{fig:time value}
\end{center}
\end{figure}

Table \ref{tab:time_value}  presents the time value given by the
difference between the GAO total value and its intrinsic value.
The volatility of an option is an important factor for this time
value since this value depends on the time until maturity and the
volatility of the underlying instrument's price. The time value
reflects the probability that the option will gain in intrinsic
value or become profitable to exercise before maturity. Figure
\ref{fig:time value} is a plot of the time value given by
 all considered models. For a deeper analysis about time value we refer the interested reader to \cite{van_Haastrecht_GAO}. \\

\begin{table}\footnotesize
\begin{center}
{
\begin{tabular}{|c|c|c|c|c|c|}
\hline
                              \multicolumn{ 6}{|c|}{{\bf GAO Time Value US}} \\
\hline
  {$g$} & {\bf BSHW } & {\bf SZHW} & {\bf LVHW1} & {\bf LVHW2} & {\bf LVHW3 } \\
\hline
       7\% &      0.9069 & 1.0460 & 1.0218&  0.9705 & 1.0593  \\
\hline
       8\% &       3.1600 & 3.5678& 3.4634& 3.2996&  3.5629  \\
\hline
    8.88\% &      7.1019 &   7.8692 &  7.5845  & 7.3323 & 7.8085  \\
\hline
       9\% &       6.4575& 7.2745&  7.0452&  6.7439&  7.2525 \\
\hline
      10\% &      2.9527&  4.2140&  3.7626&  3.3438& 4.1075 \\
\hline
      11\% &       1.0698&  2.6935&  2.3269 & 1.8009 & 2.7766 \\
\hline
      12\% &        0.4691 & 2.3658& 1.8968&  1.2865  &2.4063 \\
\hline
      13\% &       0.5152&  2.7034&  1.8755&  1.2032 & 2.4291 \\
\hline
                             \multicolumn{ 6}{|c|}{{\bf GAO Time Value EUR}} \\
\hline
  {$g$} & {\bf BSHW } & {\bf SZHW} & {\bf LVHW1} & {\bf LVHW2} & {\bf LVHW3 } \\
\hline
       7\% &       \phantom{-}0.3956&   \phantom{-}0.5831 &  \phantom{-}0.4864&   \phantom{-}0.4288  &  \phantom{-}0.5024 \\
\hline
       8\% &        \phantom{-}2.2594 &  \phantom{-}2.8827&  \phantom{-}2.6222&  \phantom{-}2.4174&   \phantom{-}2.7049 \\
\hline
   8.46\% &      \phantom{-}4.0959 &  \phantom{-}4.9816 &   \phantom{-}4.6647 &   \phantom{-}4.3232  &  \phantom{-}4.7192 \\
\hline
       9\% &        \phantom{-}1.1839&  \phantom{-}2.3673&   \phantom{-}1.9775&   \phantom{-}1.5391 &  \phantom{-}2.1672 \\
\hline
      10\% &       -1.7792 & -0.1831& -0.6030& -1.2257& -0.3195 \\
\hline
      11\% &       -2.8435 & -1.0044& -1.4098& -2.1714& -1.0884 \\
\hline
      12\% &      -3.2656 & -1.2475& -1.6625& -2.5162& -1.3139  \\
\hline
      13\% &      -3.5570 & -1.3704& -1.8134& -2.7424& -1.4376 \\
\hline
\end{tabular}
\caption{Comparison of GAO time values of the BSHW, the SZHW and
the LVHW1, LVHW2 and LVHW3 models for different guaranteed annuity
rates $g$. \label{tab:time_value}}}
\end{center}
\end{table}

These results show that the use of a non constant volatility model
such as the SZHW model and the LVHW model has a significant impact
on the total value and the time value of GAOs. Furthermore, the
term structure of the implied volatility surface has also an
influence on the total value. The numerical results show that the
LVHW3 prices tend to the SZHW prices, whereas the LVHW2 prices are
the lowest of the three LVHW models, but still above the BSHW
prices. In the LVHW1 model prices remain between BSHW and SZHW values.\\

The fact that GAO prices obtained in different implied volatility
scenarios turn out to be significantly different, underlines that
it is most important to take always into account the whole implied
volatility surface. This impact in GAO value can be justified by
equation (\ref{x_dyn_underQS}) where you can see the influence of
the level of the equity spot
$S_t$ in the dynamics of $x_t$ under the measure $Q_S$.\\

In \cite{van_Haastrecht_GAO}, it is pointed out that GAO values
are also particularly sensible to three different risk drivers
namely the survival probabilities, the fund value and the interest
rate market curve. From equation (\ref{GAO_under_QS}), we can
easily deduce that an increase of $x\%$ of the equity fund value
$S(0)$ will induce an $x\%$ increase of the GAO value. It is also
clear that a shift in the mortality table will induce a shift in
the GAO value in the same direction. Finally, a shift down applied
to the interest rates curve will increase the GAO value. The
sensibility of GAO prices with respect to the implied volatility,
the survival probabilities, the fund value and the interest rates
market curve underline the fact that for all practical purposes,
the market data used can be as important as the model used.

\subsection{GMIB Rider \label{section:GMIB_results}}

In this section, we analyze how the BSHW, the SZHW and the LVHW
models behave in the pricing of a GMIB Rider. This product has a
strong dependence on the path of the equity fund $S$ coming from
the anniversary component in the payoff, namely $\underset{n \in
A}{max}(S(n)) g \ddot{a}_{x}(T)$ (see equation \eqref{GMIB RIDER
payoff}), where $A$ is the set of anniversary dates $A= \{
1,2,...,T \}$. We use exactly the same initial data as for the
GAO. More precisely, the policyholder is assumed to be 55 years
old with a retirement age of 65 (i.e. the maturity $T$ of the GMIB
Rider is 10 years). The fund value at time $0$, $S(0)$ is assumed
to be 100. The survival rates are based on the PNMA00 table of the
Continuous Mortality Investigation (CMI) for male pensioners. We
present numerical results obtained in the US market only because
European market leads to similar behavior in pricing and
conclusions. The parameters used in each model are those obtained
after calibration as explained in Section
\ref{section_Calibration}. In Table \ref{tab:GMIB_results_0}, we
compare the price of a GMIB Rider for eight different guaranteed
annuity rates $g$ and three different guaranteed annual rates
$r_g$ (computed by using Monte Carlo simulations with 100000
simulations and 5000 steps). Note that currently, the standard
rate $r_g$ offered by insurance companies is around $5\%$ (see \cite{GMIB}).\\

\begin{figure}[h!]
\begin{center}
\includegraphics[height=4.3cm]{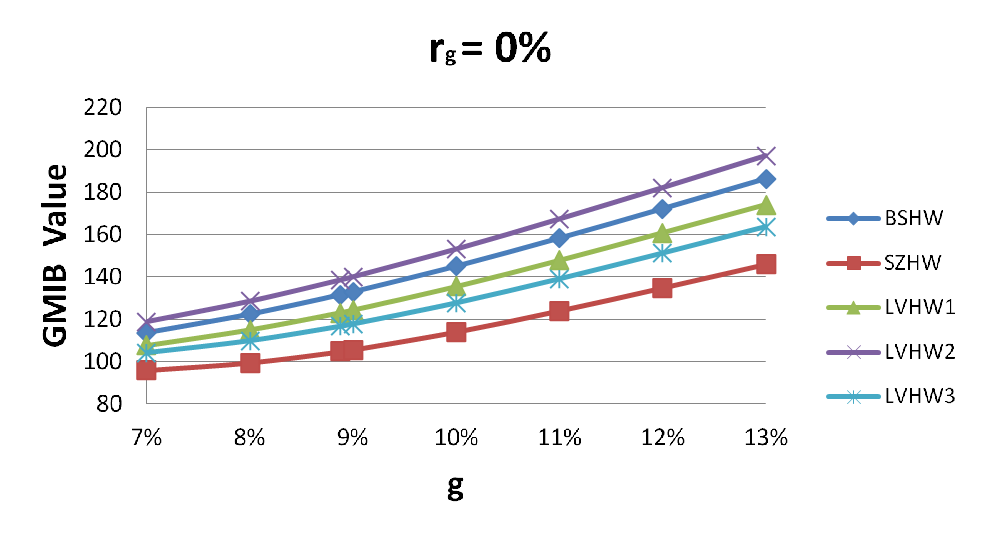}
\includegraphics[height=4.3cm]{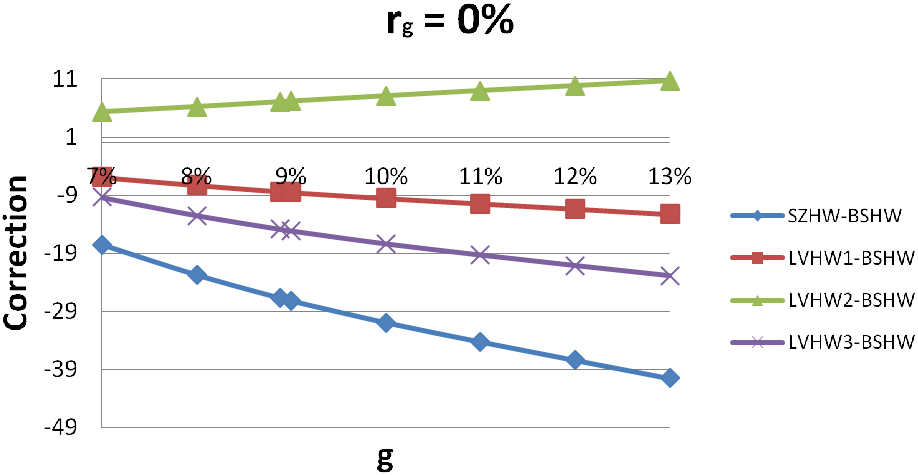}
\includegraphics[height=4.3cm]{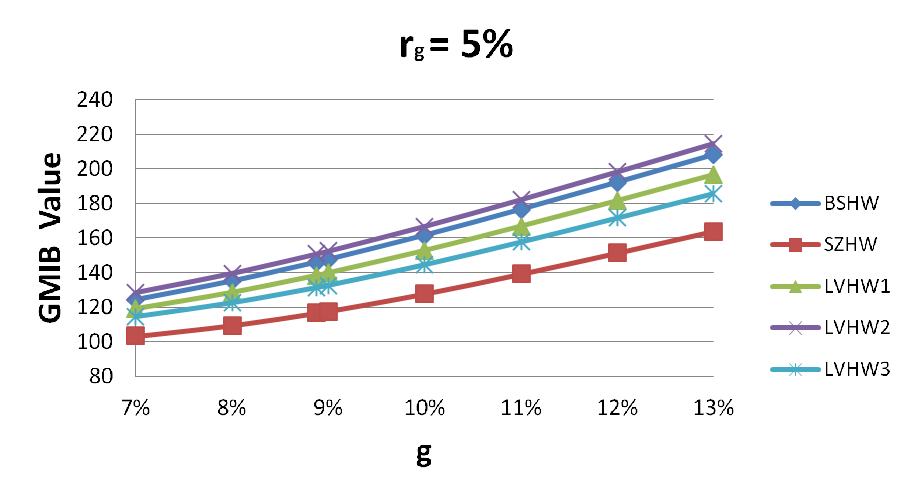}
\includegraphics[height=4.3cm]{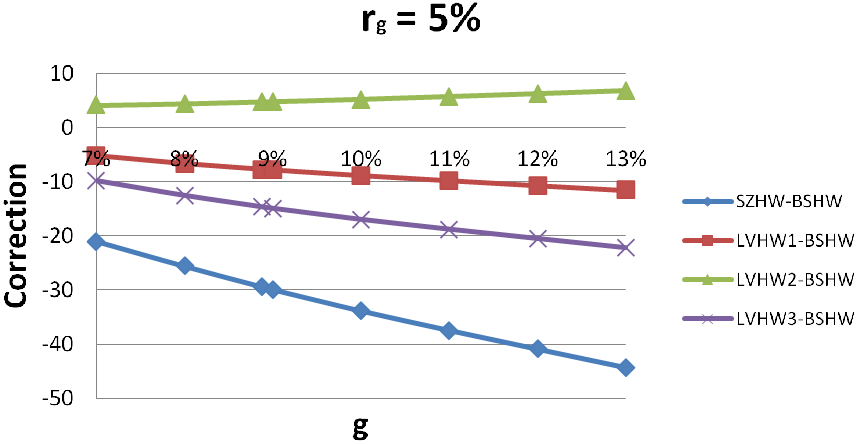}
\includegraphics[height=4.3cm]{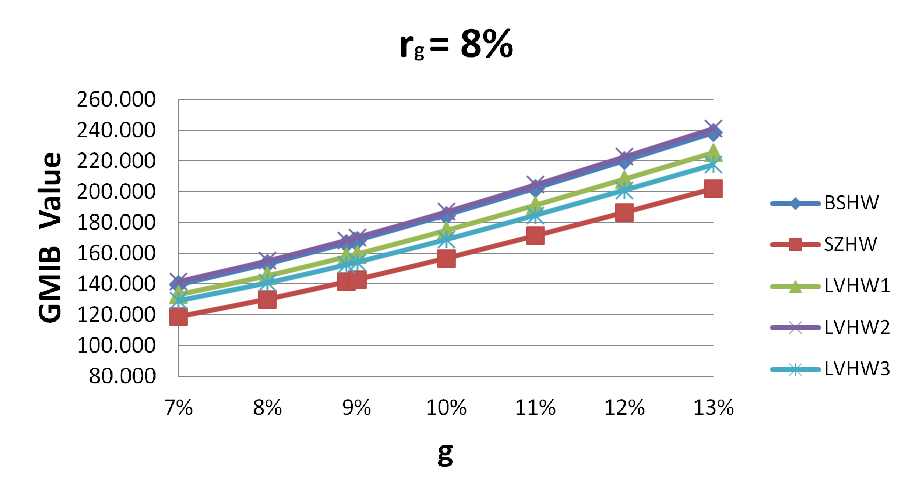}
\includegraphics[height=4.3cm]{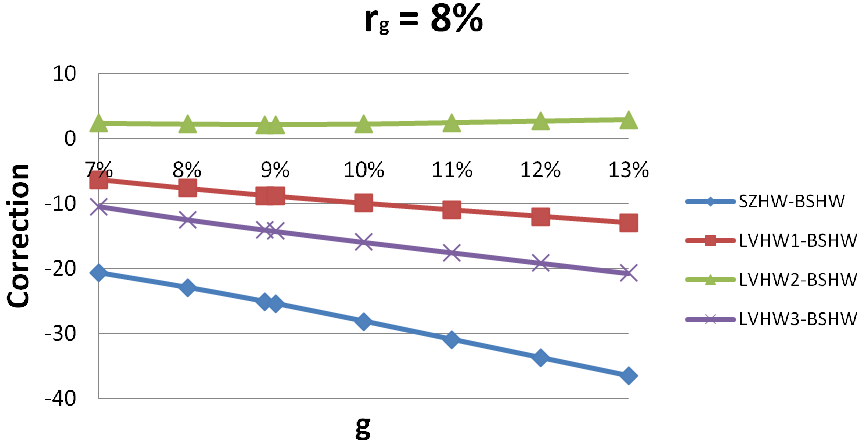}
\caption{Graphical representation of GMIB Rider total values given
by the SZHW, the LVHW1, LVHW2 and LVHW3 models for different
guaranteed annuity rates $g$ and for different guaranteed annual
rates $r_g$. The corrections given by each model with respect to
the BSHW model are also presented.}\label{fig:GMIB}
\end{center}
\end{figure}

Contrarily to the results obtained for GAOs (in Section
\ref{section GAO results}), the corrections given to GMIB values
by the SZHW model with respect to the BSHW ones are always
negative. For GMIB Riders the highest values are always given by
the LVHW2 model while the smallest values are coming from the SZHW
model. In the GAO case, the SZHW and the LVHW3 GAO values where
quite close and leading to the highest values among the observed
models. For GMIB Riders, however, the smallest prices are given by
the SZHW and the LVHW3 models, as illustrated in Figure
\ref{fig:GMIB}.

\begin{table}\footnotesize
\begin{center}
{
\begin{tabular}{|c|c|c|c|c|c|c|c|c|c|c|}
\hline
                              \multicolumn{ 11}{|c|}{{\bf   US GMIB Rider Total Value $r_g = 0 \%$ }} \\
\hline
  {$g$} & {\bf BSHW } & {\bf SE} & {\bf SZHW} & {\bf SE}  & {\bf LVHW1} & {\bf SE} & {\bf LVHW2} & {\bf SE} & {\bf LVHW3 } & {\bf SE} \\
\hline
       7\% &   113.5440 &   0.2892 &  \phantom{0}95.9552 &   0.0720  & 107.6490  &  0.1235 & 118.8630 & 0.1497 & 104.1640   & 0.1185  \\
\hline
       8\% &   122.3090  &  0.2951  &  \phantom{0}99.5474 &    0.0776 &  115.0510  &  0.1298 & 128.5280  &  0.1577 & 109.8140   & 0.1233 \\
\hline
      8.88\% & 131.4990  &  0.3063 & 104.7690  &  0.0810 & 123.1420  &  0.1375 & 138.5730  &  0.1675 & 116.6260 &   0.1300  \\
\hline
       9\% & 132.8540  &  0.3083 & 105.6330 & 0.0896  & 124.3560 &   0.1387 &  140.0450  &  0.1690 & 117.6870  & 0.1310 \\
\hline
      10\% & 144.9830  &  0.3290 & 114.0400 &   0.0900 &  135.4570  & 0.1503 &  153.1060 &   0.1830  &127.5870   & 0.1414 \\
\hline
      11\% & 158.2590 & 0.3557 & 124.0140   & 0.0938 & 147.8150   & 0.1636  & 167.3030   & 0.1990  & 138.9400   & 0.1536 \\
\hline
      12\% & 172.1850  &  0.3857 & 134.7800   & 0.1001 & 160.8320  & 0.1779  & 182.0900  &  0.2163 & 151.0450   & 0.1669 \\
\hline
      13\% & 186.3910  &  0.4171  & 145.8680   & 0.1077 & 174.1050 &   0.1925  & 197.1540 & 0.2341 & 163.4870  &  0.1806  \\
\hline
                              \multicolumn{ 11}{|c|}{{\bf   US GMIB Rider Total Value $r_g = 5 \% $ }} \\
\hline
  {$g$} & {\bf BSHW } & {\bf SE} & {\bf SZHW} & {\bf SE}  & {\bf LVHW1} & {\bf SE} & {\bf LVHW2} & {\bf SE} & {\bf LVHW3 } & {\bf SE} \\
\hline
       7\% &   124.253 &0.2770  &103.267 & 0.0713 &   119.021& 0.1325 &   128.412 & 0.1577 &   114.516 & 0.1238 \\
\hline
       8\% &   135.096 &0.2804  &109.527 &0.0738 & 128.454  & 0.1404 & 139.552 & 0.1666 & 122.620 & 0.1308 \\
\hline
      8.88\% & 145.994 &0.2897  &116.600 &0.0768& 138.313 &0.1495 &150.778 &0.1767& 131.384 &0.1392 \\
\hline
       9\% & 147.576 &0.2914  &117.684 &0.0818& 139.766 &0.1509 &   152.407 &0.1783 &132.699 &0.1405 \\
\hline
      10\% & 161.528 &0.3105 &127.717 &0.0830& 152.701 &0.1637&    166.769 &0.1927 &144.587 &0.1526\\
\hline
      11\% & 176.537 &0.3354 &139.129 &0.0859 &166.782& 0.1783 &182.269 &0.2093 &157.761 &0.1663 \\
\hline
      12\% & 192.154 &0.3637 &151.289 &0.0877 &  181.503& 0.1938 &198.384 &0.2272 &171.643 &0.1809 \\
\hline
      13\% & 208.030  &0.3934 & 163.754 &0.0944& 196.494& 0.2097&214.782 &0.2459 &185.806 &0.1959\\
\hline
                              \multicolumn{ 11}{|c|}{{\bf   US GMIB Rider Total Value $r_g = 8 \% $ }} \\
\hline
  {$g$} & {\bf BSHW } & {\bf SE} & {\bf SZHW} & {\bf SE}  & {\bf LVHW1} & {\bf SE} & {\bf LVHW2} & {\bf SE} & {\bf LVHW3 } & {\bf SE} \\
\hline
       7\% & 139.3960  &  0.2579&  118.7200   & 0.0759 & 133.1170 &   0.1441 &141.7610&    0.1600 & 128.9560 &   0.1339 \\
\hline
       8\% & 153.0990 &   0.2589& 130.1780  &  0.0775 & 145.4630 &   0.1542& 155.3320 &   0.1710 &140.6420   & 0.1439 \\
\hline
      8.88\% & 166.4050 &   0.2661& 141.3230 &   0.0820 &157.6800& 0.1651 & 168.5760 &0.1831  &152.3680 &0.1547 \\
\hline
       9\% & 168.3040 &0.2676 &142.9100 &   0.0828& 159.4420 &   0.1667 &170.4710 &0.1849 &154.0570 &0.1563 \\
\hline
      10\% & 184.7890 &0.2834 &156.6750 &0.0904 & 174.8370& 0.1812 &187.0320 &0.2013 &168.8660  &  0.1705 \\
\hline
      11\% & 202.2100 &   0.3051& 171.2920& 0.0988& 191.2250 &0.1972 & 204.6760& 0.2197 &184.6850 &0.1863 \\
\hline
      12\% & 220.1720 &0.3304 &186.4520 &0.1075 &208.1920 &0.2141& 222.8780 &0.2389 &201.0500 &0.2027\\
\hline
      13\% & 238.3830 &0.3571 &201.8650& 0.1163& 225.4030 &0.2314 &241.3100 &0.2586 &217.6730& 0.2194\\
\hline
\end{tabular}
\caption{Comparison of GMIB Rider total values of the BSHW, the
SZHW and the LVHW1, LVHW2 and LVHW3 models for different
guaranteed annuity rates $g$ and guaranteed annual rates $r_g$ of
$0 \%$, $5\%$ and of $8 \%$. \label{tab:GMIB_results_0}}}
\end{center}
\end{table}

\newpage

\subsection{Barrier GAOs \label{section: Barrier GAO results}}

In this subsection, we compare the LVHW price, the SZHW price and
the price obtained by using the three different cases of the LVHW
model for ``down-and-out GAOs" and ``down-and-in GAOs" given by
equation (\ref{DO_GAO_equation}) and (\ref{DI_GAO_equation})
respectively and computed by using Monte Carlo simulations (100
000 simulations and 5000 steps). While GMIB Riders are
path-dependent especially in $S(t)$,  these barrier options are
particularly dependent on the path of the interest rates $r(t)$.
In Table \ref{tab:BSHW_DO GAO total value}, \ref{tab:SZHW_DO GAO
total value}, \ref{tab:LVHW1_DO GAO total value},
\ref{tab:LVHW2_DO GAO total value} and \ref{tab:LVHW3_DO GAO total
value}, total values of ``down-and-out GAOs" with five different
barriers are presented. The first barrier is taken to be equal to
$B=-0.015$ and corresponds to a market annuity rate $r_x(T)$ equal
to 8\%. More precisely, when $x^*=-0.015$ and $r_x(T)=8 \%$, the
 definition of market annuity rate holds, namely,

 $$\overset{\omega - (x+T)}{\underset{n=0}\sum}
 \phantom{.}_{n} p_{x+T} A(T,T+n) e^{-
b(T,T+n) x^*}=1/r_x(T).$$

The other barriers correspond to rates $r_x(T)$ of 7\%, 6\%, 5\%
and 4\% respectively or equivalently to barrier levels $B$ equal
to -0.033, -0.05225, -0.0791 and -0.1019 respectively. Note that
since the initial value of $x(0)$ is equal to 0, the barrier level
$B$ has to be smaller and consequently strictly negative. The
value of a ``pure GAO" is given in the last column of Table
\ref{tab:BSHW_DO GAO total value}, \ref{tab:SZHW_DO GAO total
value}, \ref{tab:LVHW1_DO GAO total value}, \ref{tab:LVHW2_DO GAO
total value} and \ref{tab:LVHW3_DO GAO total value}. When the
barrier level is equal to 4\%, the ``down-and-out GAO" value is
close to the GAO value. In that case, the survival probability of
the ``down-and-out GAO" is close to one. A graphical
representation of the corrections given by each model with respect
to the BSHW model can be found in Figure \ref{fig:DO
GAO corrections}, and this for eight different guaranteed annuity rates $g$.\\

\begin{figure}
\begin{center}
\includegraphics[height=3.5cm]{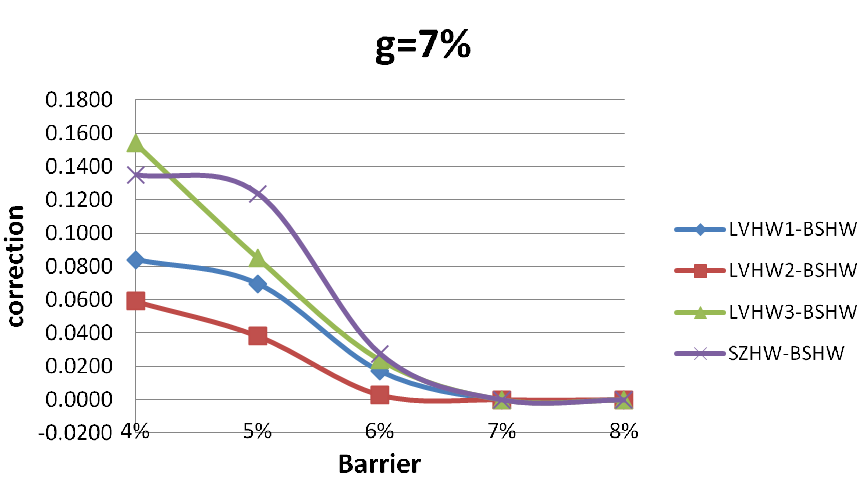}
\includegraphics[height=3.5cm]{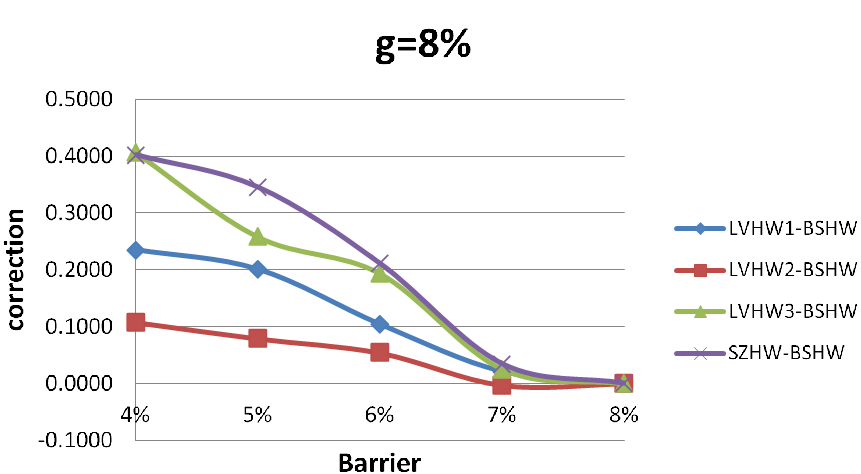}
\includegraphics[height=3.5cm]{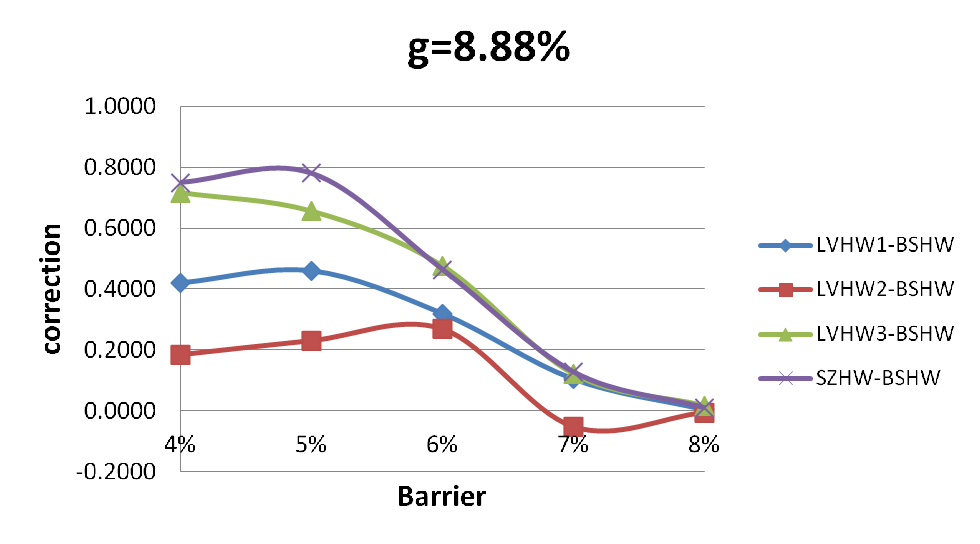}
\includegraphics[height=3.5cm]{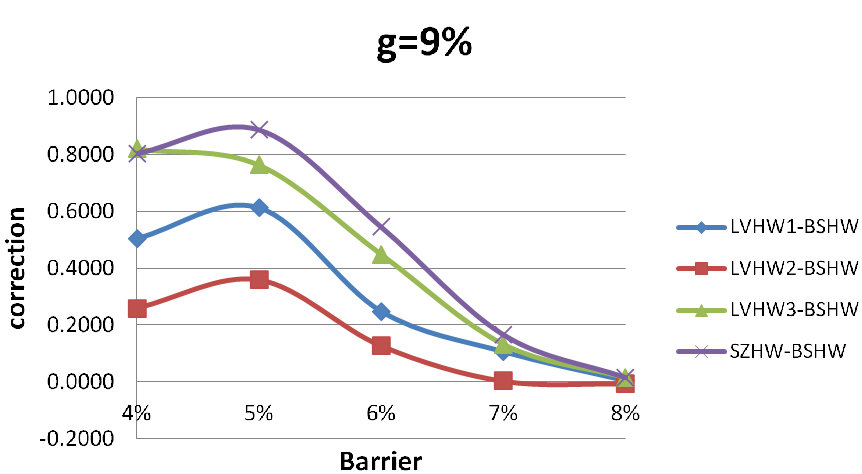}
\includegraphics[height=3.5cm]{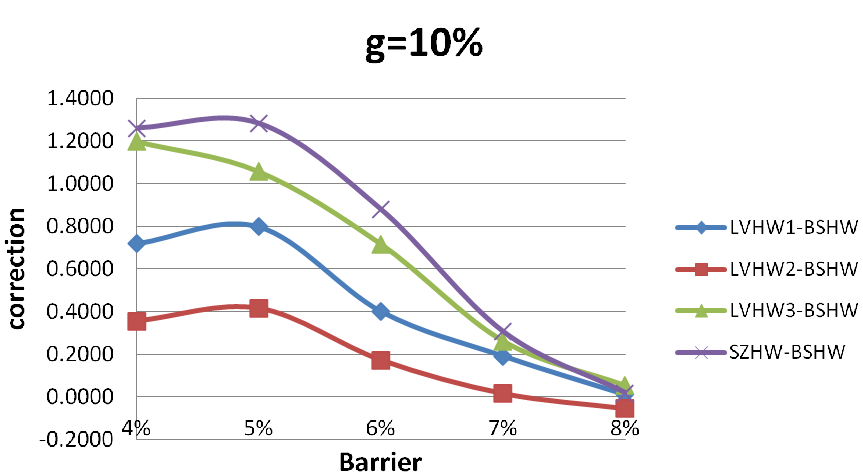}
\includegraphics[height=3.5cm]{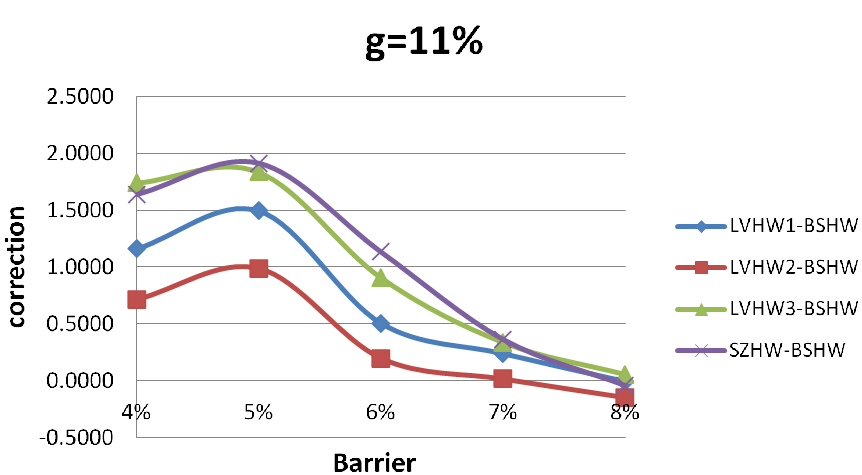}
\includegraphics[height=3.5cm]{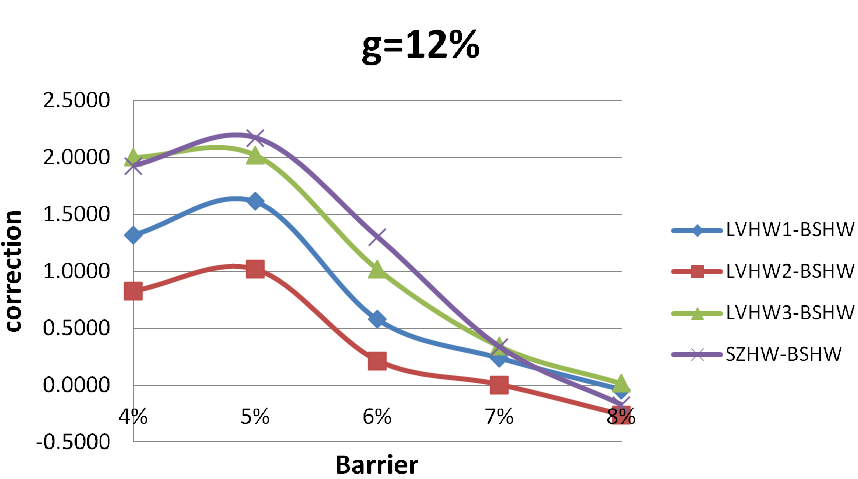}
\includegraphics[height=3.5cm]{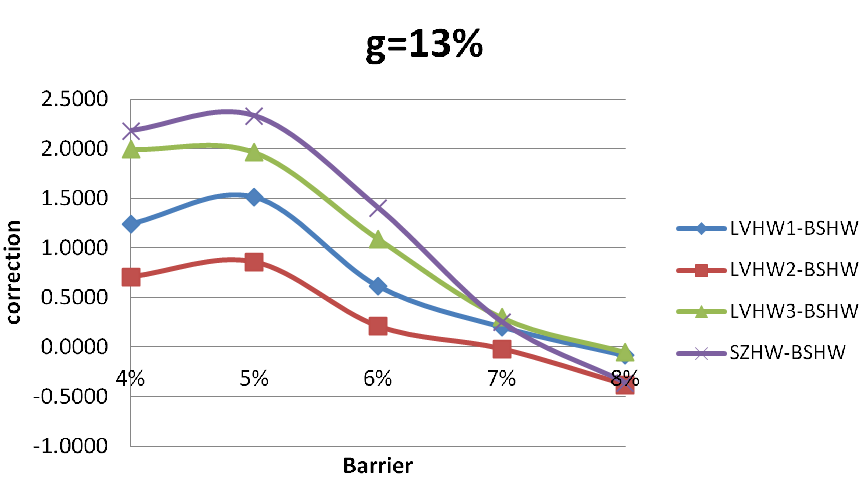}
\caption{Graphical representation of DO GAO corrections of the
SZHW, the LVHW1, LVHW2 and LVHW3 models for different guaranteed
annuity rates $g$ with respect to the BSHW model.}\label{fig:DO
GAO corrections}
\end{center}
\end{figure}

\begin{table}\footnotesize
\begin{center}{
\begin{tabular}{|c|c|c|c|c|c|c|}
\hline
                     \multicolumn{ 6}{|c|}{{\bf BSHW DO GAO Total Value (US)  }} & {\bf BSHW GAO } \\
\hline
    \backslashbox{$g$}{Barrier} &       8.00\% &      7.00\% &       6.00\% &       5.00\% &       4.00\% &            \\
\hline
    7\% &     \phantom{0}0.0000 &     \phantom{0}0.0000 &     \phantom{0}0.2314 &     \phantom{0}0.8138 &     \phantom{0}0.9046 &     \phantom{0}0.9069 \\
    (SE)   &   \phantom{0}(0.0000)  &   \phantom{0}(0.0000)   &   \phantom{0}(0.0041)  & \phantom{0}(0.0112) &  \phantom{0}(0.0131) &    \\
    \hline
    8\% &     \phantom{0}0.0000 &     \phantom{0}0.3411 &     \phantom{0}1.7787 &     \phantom{0}3.0548 &     \phantom{0}3.1415 &     \phantom{0}3.1600 \\
    (SE)   &   \phantom{0}(0.0000) & \phantom{0}(0.0047) & \phantom{0}(0.0150)& \phantom{0}(0.0238)&  \phantom{0}(0.0256) &      \\
    \hline
    8.88\% &     \phantom{0}0.1607 &  \phantom{0}2.0274 & \phantom{0}5.0804 & \phantom{0}6.8923 & \phantom{0}7.0821 & \phantom{0}7.1019 \\
    (SE)   &  \phantom{0}(0.0028) &  \phantom{0}(0.0144) & \phantom{0}(0.0275) & \phantom{0}(0.0369) & \phantom{0}(0.0387)  &     \\
    \hline
    9\% &     \phantom{0}0.2273 &     \phantom{0}2.3625 &     \phantom{0}5.6478 &     \phantom{0}7.6506 &     \phantom{0}7.7299 &     \phantom{0}7.7384 \\
    (SE)   &    \phantom{0}(0.0036) & \phantom{0}(0.0159) &  \phantom{0}(0.0293) & \phantom{0}(0.0387) & \phantom{0}(0.0405) &     \\
    \hline
   10\% &     \phantom{0}1.5112 &     \phantom{0}6.6738 &    12.0277 &    14.7523 &    14.8721 &    14.8802 \\
   (SE)   &   \phantom{0}(0.0125) & \phantom{0}(0.0296) & \phantom{0}(0.0435) &  \phantom{0}(0.0529) & \phantom{0}(0.0546) &      \\
   \hline
   11\% &     \phantom{0}4.0316 &    12.8157 &    20.2984 &    23.0443 &    23.5709 &    23.6438 \\
   (SE)   &   \phantom{0}(0.0240) & \phantom{0}(0.0429) & \phantom{0}(0.0554)&  \phantom{0}(0.0643)&  \phantom{0}(0.0661) &     \\
   \hline
   12\% &     \phantom{0}7.3759 &    20.0008 &    29.6144 &    33.0740 &    33.6186 &    33.6896 \\
   (SE)   &   \phantom{0}(0.0365) & \phantom{0}(0.0553) & \phantom{0}(0.0650) &  \phantom{0}(0.0731) &  \phantom{0}(0.0749) &     \\
   \hline
   13\% &    11.0977 &    27.6196 &    39.3708 &    44.2385 &    44.3125 &    44.3822 \\
   (SE)   &   \phantom{0}(0.0497)&  \phantom{0}(0.0673)&  \phantom{0}(0.0733) & \phantom{0}(0.0803) & \phantom{0}(0.0821) &     \\
\hline
\end{tabular}
\caption{``Down-and-out GAO" total values given by the BSHW model
for eight different guaranteed annuity rates $g$ and for five
different barriers. The pure GAO values for the eight different
guaranteed annuity rates $g$ are also given in the last column.
\label{tab:BSHW_DO GAO total value}}}
\end{center}
\end{table}

\begin{table}\footnotesize
\begin{center}{
\begin{tabular}{|c|c|c|c|c|c|c|}
\hline
                     \multicolumn{ 6}{|c|}{{\bf SZHW DO GAO Total Value (US)  }} & {\bf SZHW GAO } \\
\hline
    \backslashbox{$g$}{Barrier} &       8.00\% &      7.00\% &       6.00\% &       5.00\% &       4.00\% &            \\
\hline
    7.00\% &     \phantom{0}0.0000 &     \phantom{0}0.0000 &     \phantom{0}0.2591 &     \phantom{0}0.9375 &     \phantom{0}1.0345 &     \phantom{0}1.0460 \\
    (SE)   &   \phantom{0}(0.0000)&  \phantom{0}(0.0000)&  \phantom{0}(0.0043) & \phantom{0}(0.0122) & \phantom{0}(0.0141) &     \\
    \hline
    8.00\% &     \phantom{0}0.0000 &     \phantom{0}0.3753 &     \phantom{0}1.9895 &     \phantom{0}3.3998 &     \phantom{0}3.5433 &     \phantom{0}3.5678 \\
    (SE)   &   \phantom{0}(0.0000) & \phantom{0}(0.0049) & \phantom{0}(0.0156)& \phantom{0}(0.0253)&  \phantom{0}(0.0272)&     \\
    \hline
    8.88\% &     \phantom{0}0.1704 &  \phantom{0}2.1545 & \phantom{0}5.5442 & \phantom{0}7.6723 & \phantom{0}7.8320 & \phantom{0}7.8692 \\
    (SE)   &   \phantom{0}(0.0029) & \phantom{0}(0.0149)&  \phantom{0}(0.0285) & \phantom{0}(0.0387)&  \phantom{0}(0.0405) &     \\
    \hline
    9.00\% &     \phantom{0}0.2413 &     \phantom{0}2.5278 &     \phantom{0}6.1930 &     \phantom{0}8.3709 &     \phantom{0}8.5336 &     \phantom{0}8.5554 \\
    (SE)   &  \phantom{0}(0.0037)& \phantom{0}(0.0165) & \phantom{0}(0.0303) & \phantom{0}(0.0405)&  \phantom{0}(0.0423) &     \\
    \hline
   10.00\% &     \phantom{0}1.5298 &     \phantom{0}6.9793 &    12.9071 &    15.8655 &    16.0528 &    16.1414 \\
   (SE)   &  \phantom{0}(0.0127) & \phantom{0}(0.0303)&  \phantom{0}(0.0445)&  \phantom{0}(0.0546) & \phantom{0}(0.0563) &     \\
   \hline
   11.00\% &     \phantom{0}3.9929 &    13.1818 &    21.4323 &    25.1681 &    25.1995 &    25.2674 \\
   (SE)   &   \phantom{0}(0.0242)&  \phantom{0}(0.0437)&  \phantom{0}(0.0562)&  \phantom{0}(0.0657) & \phantom{0}(0.0675) &     \\
    \hline
   12.00\% &     \phantom{0}7.2008 &    20.3345 &    30.9134 &    35.4296 &    35.5052 &    35.5863 \\
   (SE)   &  \phantom{0}(0.0368) & \phantom{0}(0.0564)& \phantom{0}(0.0658)&  \phantom{0}(0.0743)&  \phantom{0}(0.0761) &     \\
    \hline
   13.00\% &    10.7340 &    27.8709 &    40.7734 &    46.0700 &    46.3932 &    46.5705 \\
   (SE)   &   \phantom{0}(0.0500)&  \phantom{0}(0.0689)&  \phantom{0}(0.0742)&  \phantom{0}(0.0815)&  \phantom{0}(0.0833)&      \\
\hline
\end{tabular}
\caption{``Down-and-out GAO" total values given by the SZHW model
for eight different guaranteed annuity rates $g$ and for five
different barriers. The pure GAO values for the eight different
guaranteed annuity rates $g$ are also given in the last column.
\label{tab:SZHW_DO GAO total value}}}
\end{center}
\end{table}

\begin{table}\footnotesize
\begin{center}{
\begin{tabular}{|c|c|c|c|c|c|c|}
\hline
                     \multicolumn{ 6}{|c|}{{\bf LVHW1 DO GAO Total Value (US)  }} & {\bf LVHW1 GAO } \\
\hline
    \backslashbox{$g$}{Barrier} &       8.00\% &      7.00\% &       6.00\% &       5.00\% &       4.00\% &            \\
\hline
7\% &     \phantom{0}0.0000 &     \phantom{0}0.0000 &     \phantom{0}0.2486 &     \phantom{0}0.8834 &     \phantom{0}0.9887 &     \phantom{0}1.0218 \\
(SE)   &   \phantom{0}(0.0000) & \phantom{0}(0.0000) & \phantom{0}(0.0079) & \phantom{0}(0.0133)  & \phantom{0}(0.0141) &    \phantom{0}(0.0137)  \\
    \hline
    8\% &     \phantom{0}0.0000 &     \phantom{0}0.3626 &     \phantom{0}1.8830 &     \phantom{0}3.2555 &     \phantom{0}3.3862 &     \phantom{0}3.4634 \\
    (SE)   &   \phantom{0}(0.0000) & \phantom{0}(0.0102) & \phantom{0}(0.0197)& \phantom{0}(0.0262) &  \phantom{0}(0.0270)      &   \phantom{0}(0.0265)   \\
    \hline
    8.88\% &    \phantom{0}0.1666 & \phantom{0}2.1331  & \phantom{0}5.4004 & \phantom{0}7.3525 & \phantom{0}7.5028 & \phantom{0}7.5845 \\
    (SE)   &    \phantom{0}(0.0085)  & \phantom{0}(0.0213) & \phantom{0}(0.0327) & \phantom{0}(0.0395)  & \phantom{0}(0.0403)     &  \phantom{0}(0.0398)    \\
    \hline
    9\% &     \phantom{0}0.2311 &     \phantom{0}2.4683 &     \phantom{0}5.8958 &     \phantom{0}8.0631 &     \phantom{0}8.2038 &     \phantom{0}8.3261 \\
    (SE)   &     \phantom{0}(0.0096) & \phantom{0}(0.0230) &  \phantom{0}(0.0345)&  \phantom{0}(0.0414) &  \phantom{0}(0.0421)   &    \phantom{0}(0.0416)  \\
    \hline
   10\% &     \phantom{0}1.5206 &     \phantom{0}6.8641 &    12.4280 &    15.3804 &    15.5415 &    15.6900 \\
   (SE)   &    \phantom{0}(0.0208)  & \phantom{0}(0.0370) & \phantom{0}(0.0487) & \phantom{0}(0.0556) & \phantom{0}(0.0562)      &  \phantom{0}(0.0556)    \\
    \hline
   11\% &     \phantom{0}4.0272 &    13.0565 &    20.8054 &    24.5382 &    24.7208 &    24.9009 \\
   (SE)   &    \phantom{0}(0.0336) & \phantom{0}(0.0499) & \phantom{0}(0.0604) & \phantom{0}(0.0671) & \phantom{0}(0.0677)      &  \phantom{0}(0.0670)    \\
    \hline
   12\% &     \phantom{0}7.3354 &    20.2375 &    30.1895 &    34.6890 &    34.8983 &    35.1173 \\
   (SE)   &     \phantom{0}(0.0470)  & \phantom{0}(0.0615) & \phantom{0}(0.0696) & \phantom{0}(0.0759) & \phantom{0}(0.0765)     &  \phantom{0}(0.0757)    \\
    \hline
   13\% &    11.0113 &    27.8210 &    39.9830 &    45.2484 &    45.4933 &    45.7425 \\
   (SE)   &    \phantom{0}(0.0608) & \phantom{0}(0.0724)  & \phantom{0}(0.0775) & \phantom{0}(0.0833) & \phantom{0}(0.0839)   &  \phantom{0}(0.0830)    \\
 \hline
\end{tabular}
\caption{``Down-and-out GAO" total values given by the LVHW1 model
for eight different guaranteed annuity rates $g$ and for five
different barriers. In the last column, one has the pure GAO
values for the eight different guaranteed annuity rates $g$.
\label{tab:LVHW1_DO GAO total value}}}
\end{center}
\end{table}

\begin{table}\footnotesize
\begin{center}{
\begin{tabular}{|c|c|c|c|c|c|c|}
\hline
 \multicolumn{ 6}{|c|}{{\bf LVHW2 DO GAO Total Value (US)  }} & {\bf LVHW2 GAO } \\
 \hline
    \backslashbox{$g$}{Barrier} &       8.00\% &      7.00\% &       6.00\% &       5.00\% &       4.00\% &            \\
\hline
    7\% &     \phantom{0}0.0000 &     \phantom{0}0.0000 &     \phantom{0}0.2343 &     \phantom{0}0.8521 &   \phantom{0}0.9537 &     \phantom{0}0.9705 \\
    (SE)   &   \phantom{0}(0.0000) & \phantom{0}(0.0000) & \phantom{0}(0.0077)&  \phantom{0}(0.0131)&  \phantom{0}(0.0139)      &   \phantom{0}(0.0134)   \\
    \hline
    8\% &     \phantom{0}0.0000 &     \phantom{0}0.3384 &     \phantom{0}1.8329 &     \phantom{0}3.1339 &    \phantom{0}3.2095 &     \phantom{0}3.2996 \\
    (SE)   &    \phantom{0}(0.0000) & \phantom{0}(0.0101) & \phantom{0}(0.0195) & \phantom{0}(0.0259)&  \phantom{0}(0.0267)       &  \phantom{0}(0.0259)    \\
    \hline
    8.88\% &     \phantom{0}0.1554 &  \phantom{0}1.9745 & \phantom{0}5.3501  & \phantom{0}7.1236&  \phantom{0}7.2671 & \phantom{0}7.3323 \\
    (SE)   &   \phantom{0}(0.0081) & \phantom{0}(0.0211) & \phantom{0}(0.0324)&  \phantom{0}(0.0392) & \phantom{0}(0.0400)     &    \phantom{0}(0.0390)  \\
    \hline
    9\% &     \phantom{0}0.2206 &     \phantom{0}2.3656 &     \phantom{0}5.7748 &     \phantom{0}7.8091 &    \phantom{0}7.9390 &     \phantom{0}8.0248 \\
    (SE)   &     \phantom{0}(0.0092) & \phantom{0}(0.0228) & \phantom{0}(0.0342)& \phantom{0}(0.0410) & \phantom{0}(0.0418)      &    \phantom{0}(0.0408)    \\
    \hline
   10\% &     \phantom{0}1.4562 &     \phantom{0}6.6893 &    12.2000 &    14.9976 &    15.1776 &    15.2712 \\
   (SE)   &     \phantom{0}(0.0203) & \phantom{0}(0.0368) & \phantom{0}(0.0484) & \phantom{0}(0.0552)& \phantom{0}(0.0560)     &    \phantom{0}(0.0549)  \\
    \hline
   11\% &     \phantom{0}3.8866 &    12.8339 &    20.4983 &    24.0320 &    24.2953 &    24.3749 \\
   (SE)   &     \phantom{0}(0.0331) & \phantom{0}(0.0497)& \phantom{0}(0.0602) & \phantom{0}(0.0667) & \phantom{0}(0.0675)       &  \phantom{0}(0.0663)    \\
    \hline
   12\% &     \phantom{0}7.1146 &    20.0048 &    29.8277 &    34.0920 &    34.4322 &    34.5070 \\
   (SE)   &  \phantom{0}(0.0465) & \phantom{0}(0.0613) & \phantom{0}(0.0695)&  \phantom{0}(0.0756)& \phantom{0}(0.0763)  &    \phantom{0}(0.0750)  \\
    \hline
   13\% &    10.7150 &    27.5991 &    39.5786 &    44.5937 &    44.9869 &    45.0702 \\
   (SE)   &    \phantom{0}(0.0602) & \phantom{0}(0.0723)& \phantom{0}(0.0774)&  \phantom{0}(0.0829) & \phantom{0}(0.0837)      &   \phantom{0}(0.0822)   \\
   \hline
\end{tabular}
\caption{``Down-and-out GAO" total values given by the LVHW2 model
for eight different guaranteed annuity rates $g$ and for five
different barriers. In the last column, one has the pure GAO
values for the eight different guaranteed annuity rates $g$.
\label{tab:LVHW2_DO GAO total value}}}
\end{center}
\end{table}

\begin{table}\footnotesize
\begin{center}{
\begin{tabular}{|c|c|c|c|c|c|c|}
\hline
\multicolumn{ 6}{|c|}{{\bf LVHW3 DO GAO Total Value (US)  }} & {\bf LVHW3 GAO } \\
\hline
    \backslashbox{$g$}{Barrier} &       8.00\% &      7.00\% &       6.00\% &       5.00\% &       4.00\% &            \\
\hline
   7\% &     \phantom{0}0.0000 &     \phantom{0}0.0000 &     \phantom{0}0.2554 &     \phantom{0}0.8988 &     \phantom{0}1.0565 &     \phantom{0}1.0593 \\
   (SE)   &  \phantom{0}(0.0000) & \phantom{0}(0.0000) & \phantom{0}(0.0081) & \phantom{0}(0.0136) & \phantom{0}(0.0144)      &   \phantom{0}(0.0142)   \\
    \hline
    8\% &     \phantom{0}0.0000 &     \phantom{0}0.3665 &     \phantom{0}1.9723 &     \phantom{0}3.3129 &     \phantom{0}3.5442 &     \phantom{0}3.5629 \\
    (SE)   &  \phantom{0}(0.0000) & \phantom{0}(0.0105) & \phantom{0}(0.0201)& \phantom{0}(0.0266)&  \phantom{0}(0.0274)      & \phantom{0}(0.0271)     \\
    \hline
    8.88\% &     \phantom{0}0.1784  & \phantom{0}2.1492 & \phantom{0}5.5560 & \phantom{0}7.5477 & \phantom{0}7.7981 & \phantom{0}7.8085 \\
    (SE)   &    \phantom{0}(0.0086) & \phantom{0}(0.0216) & \phantom{0}(0.0332) & \phantom{0}(0.0399)  & \phantom{0}(0.0407)      &  \phantom{0}(0.0404)    \\
    \hline
    9\% &     \phantom{0}0.2411 &     \phantom{0}2.4941 &     \phantom{0}6.0965 &     \phantom{0}8.2142 &     \phantom{0}8.5003 &     \phantom{0}8.5334 \\
    (SE)   &   \phantom{0}(0.0098) & \phantom{0}(0.0233) & \phantom{0}(0.0350) & \phantom{0}(0.0417) & \phantom{0}(0.0425)     &   \phantom{0}(0.0423)   \\
    \hline
   10\% &     \phantom{0}1.5628 &     \phantom{0}6.9347 &    12.7441 &    15.6396 &    16.0209 &    16.0349 \\
   (SE)   &    \phantom{0}(0.0210) & \phantom{0}(0.0373) & \phantom{0}(0.0492)&  \phantom{0}(0.0559) & \phantom{0}(0.0567) &    \phantom{0}(0.0563)    \\
    \hline
   11\% &     \phantom{0}4.0876 &    13.1534 &    21.2053 &    24.8806 &    25.3019 &    25.3506 \\
   (SE)   &    \phantom{0}(0.0340)  & \phantom{0}(0.0502) & \phantom{0}(0.0609) & \phantom{0}(0.0674)  & \phantom{0}(0.0682)     & \phantom{0}(0.0676)     \\
    \hline
   12\% &     7.3917 &    20.3405 &    30.6314 &    35.0938 &    35.5751 &    35.6268 \\
   (SE)   &  \phantom{0}(0.0474)  & \phantom{0}(0.0617)& \phantom{0}(0.0701) & \phantom{0}(0.0763) & \phantom{0}(0.0771)     &  \phantom{0}(0.0763)   \\
    \hline
   13\% &    11.0452 &    27.9176 &    40.4625 &    45.7019 &    46.2445 &    46.2961 \\
   (SE)   &   \phantom{0}(0.0612) & \phantom{0}(0.0727) & \phantom{0}(0.0780) & \phantom{0}(0.0837) & \phantom{0}(0.0845)   &   \phantom{0}(0.0836)   \\
    \hline
\end{tabular}
\caption{``Down-and-out GAO" total values given by the LVHW3 model
for eight different guaranteed annuity rates $g$ and for five
different barriers. In the last column, one has the pure GAO
values for the eight different guaranteed annuity rates $g$.
\label{tab:LVHW3_DO GAO total value}}}
\end{center}
\end{table}

The price of the ``down-and-in GAO" can easily be computed from
the price of the ``down-and-out GAO" and the ``pure GAO" by using
the relation given by equation (\ref{DI_GAO_equation}). In
Appendix \ref{DIGAO_results}, Table \ref{tab1:DI GAO total value},
total values of US ``down-and-in GAOs" for the eight different
guaranteed annuity rates $g$ and for the five different barriers
are presented. Figure \ref{fig:DI GAO_corrections} in Appendix
\ref{DIGAO_results} illustrates the corrections for ``down-and-in
GAOs" given by each model with respect to
the BSHW model for eight different guaranteed annuity rates $g$.\\

In the case of barrier GAOs the corrections given by each model
are more complicated to analyze in the sense that there is no
general conclusion with respect to the correction behavior. More
precisely, we are not able to answer the question which model
gives the highest values or the smallest ones because it depends
on both the barrier level and of the guaranteed annuity rate $g$.

\section{Conclusion \label{section Conclusion}}

The local volatility model with stochastic interest rates is a
suitable model to price and hedge long maturities life insurance
contracts. This model takes into account the stochastic behavior
of the interest rates as well as the vanilla market smile effects.
The local volatility captures the whole implied volatility surface
and is a deterministic function presenting an advantage for
hedging strategies in comparison with
stochastic volatility models for which the market is incomplete.\\

A first contribution of the paper is the calibration of a local
volatility surface in a stochastic interest rates framework. We
have developed a Monte Carlo approach for the calibration and this
method has successfully been tested on US and European market call
data.\\

The second contribution is the analysis of the impact of using a
local volatility model to the price of long-dated insurance
products as Variable Annuity Guarantees. More precisely, we have
compared prices of GAO, GMIB Rider and barrier GAO obtained by
using the local volatility model with stochastic interest rates to
the prices given by a constant volatility and a stochastic
volatility model all calibrated to the same data. The
particularity of the GMIB Rider is the strong dependence on the
path of the equity fund; whereas, the interest rate barrier type
options have a strong dependence on the path of interest rates.
The results confirm that calibrating such models to the vanilla
market is by no means a guarantee that derivatives will be priced
identically.\\

Where \cite{van_Haastrecht_GAO} already pointed out that using a
non constant volatility has a significant impact on the price of
GAO we generalized this conclusion to GMIB Rider and Barrier GAOs
and used local volatility models. Furthermore, we confirm that
when using market data given in \cite{van_Haastrecht_GAO}, the
constant volatility Black Scholes model with stochastic interest
rates turns out to underestimate the value of GAO compared to the
Sch\"{o}bel and Zhu stochastic volatility model with stochastic
interest rates. However, we show that in a local volatility
framework with stochastic interest rates, the price of GAO depends
on the whole option's implied volatility surface. Moreover, for
GMIB Riders, the conclusion is the opposite, the SZHW model prices
are always smaller than the corresponding BSHW model prices.\\

This paper underlines the fact that due to the sensibilities of
Variable Annuity Guarantee prices with respect to the model used
(after calibration to the Vanilla market), and also the
sensibilities with respect to data, namely, the survival
probability table and the yield curve, practitioners should be
careful in their model choice as well as
the market data chosen and the calibration of the model.\\

The results presented in this paper show that stochastic and local
volatility models, perfectly calibrated on the same market implied
volatility surface, do not imply  the same prices for Variable
Annuity Guarantees. In \cite{lipton3}, the authors underline that
the market dynamics could be better approximated by a hybrid
volatility model that contains both stochastic volatility dynamics
and local volatility ones. The study of a pure local volatility
model is crucial for the calibration of such hybrid volatility
models (see \citep{Deelstra-Rayee}). For future research we plan
to calibrate hybrid volatility models, based on the results
obtained for the pure local volatility model. We further will
study the impact of the hybrid volatility models to GAOs, GMIB
Riders and barrier GAOs. The hedging performance of all these
models is also left for future work.

\section*{Acknowledgments}
We would like to thank A. van Haastrecht, R. Plat and A. Pelsser
for providing us the Hull and White parameters and interest rate
curve data they have used in \cite{van_Haastrecht_GAO}.

{\normalsize
\bibliographystyle{plain}
\bibliography{bibliography}
}

\appendix


\section{Explicit formula for the GAO price in the BSHW and SZHW models \label{Appendix_closed_form}}

In this appendix, we recall the explicit formula for a GAO price
in the BSHW  and in the SZHW models derived in \cite{Ballotta} and
\cite{van_Haastrecht_GAO} respectively.\\

In the SZHW model, the fund value $S$, the interest rate $r$ and
the volatility are governed by the following dynamics:

\begin{eqnarray}
dS(t) &=& (r(t)-q)  S(t)  dt + \nu(t) S(t) dW^Q_{S}(t), \nonumber \\
dr(t) &=& (\theta(t) - \alpha  r(t))  dt + \sigma_{r} dW^Q_{r}(t), \nonumber \\
d\nu(t) &=& \kappa (\psi - \nu(t)) dt + \tau dW^Q_{\nu}(t).
\nonumber
\end{eqnarray}

The dynamics of the fund $S(t)$, the interest rates $r(t)$ and the
volatility of the fund $\nu(t)$ are linked by the following
correlation structure:

\begin{eqnarray}
E^{Q}[dW^Q_{S} dW^Q_{r}] = \rho_{Sr}  dt, \nonumber \\
E^{Q}[dW^Q_{S} dW^Q_{\nu}] = \rho_{S \nu} dt, \nonumber \\
E^{Q}[dW^Q_{\nu} dW^Q_{r}] = \rho_{\nu r}  dt. \nonumber
\end{eqnarray}

The explicit formula of the GAO price is given by

\begin{eqnarray}
 C(x,0,T) = \phantom{.}_{T} p_{x} g S(0)
\sum_{n=0}^{\omega - (x+T)} \phantom{.}_{n} p_{x+T} (F_n
\mathcal{N}(d_1^n) - K_n \mathcal{N}(d_2^n)),
\end{eqnarray}

\noindent where the strikes are given by $K_n = A(T,T+n) e^{-
b(T,T+n) x^* }$ with $x^*$ solving
$$\overset{\omega - (x+T)}{\underset{i=0}\sum}
 \phantom{.}_{n} p_{x+T} A(T,T+n) e^{-
b(T,T+n) x^*}=1/g$$

\noindent with

\begin{eqnarray}
 F_n &=& e^{M_n + \frac{1}{2} V_n } \nonumber \\
d_1^n &=& \frac{\ln(F_n/K_n) + \frac{1}{2} V_n }{\sqrt{V_n}}    \nonumber \\
d_2^n &=& d_1^n -  \sqrt{V_n} \nonumber \\
M_n &=& \ln(A(T,T+n)) - b(T,T+n) \mu_x \nonumber \\
V_n &=& b^2(T,T+n) \sigma^2_x
\end{eqnarray}

In the SZHW model, the mean $\mu_x$  and variance $\sigma_x$ are
given by

\begin{eqnarray}
 \mu_x &=& \rho_{Sr} \sigma_r (\frac{ \widetilde{\psi}}{\alpha} (1-e^{-\alpha T}) + \frac{ \nu(0) - \widetilde{\psi}}{\alpha - \widetilde{\kappa}} (e^{-\widetilde{\kappa} T}-e^{-\alpha T}) ) \nonumber \\
\sigma^2_x &=& \sigma^2_1 + \sigma^2_2 + 2 \rho_{12} \sigma_1
\sigma_2
\end{eqnarray}

\noindent with

\begin{eqnarray}
 \sigma_1 &=& \sigma_r \sqrt{\frac{ 1 }{2\alpha} (1-e^{-2 \alpha T}) }\nonumber \\
\sigma_2 &=&  \frac{\rho_{Sr} \sigma_r \tau}{\alpha - \widetilde{\kappa} } \sqrt{\frac{1}{2 \widetilde{\kappa}} + \frac{1}{2 \alpha} - \frac{2}{\alpha + \widetilde{\kappa}} - \frac{e^{-2 \widetilde{\kappa} T}}{2 \widetilde{\kappa}} - \frac{e^{-2 \alpha T}}{2 \alpha} + \frac{2 e^{-2(\alpha + \widetilde{\kappa}) T}}{\alpha + \widetilde{\kappa}}} \nonumber \\
\rho_{12} &=&\rho_{r\nu} \frac{\rho_{Sr} \sigma^2_r \tau}{\sigma_1
\sigma_2 (\alpha - \widetilde{\kappa}) } (\frac{1- e^{-(\alpha +
\widetilde{\kappa}) T}}{\alpha + \widetilde{\kappa}} -
\frac{1-e^{-2 \alpha T}}{2 \alpha})
\end{eqnarray}

\begin{eqnarray}
 \widetilde{\psi} &=& \frac{\psi \kappa}{\widetilde{\kappa}} \nonumber \\
\widetilde{\kappa} &=& \kappa - \rho_{S \nu} \tau\nonumber
\end{eqnarray}

The BSHW model is given by the following dynamics:

\begin{eqnarray}
dS(t) &=& (r(t)-q)  S(t)  dt + \sigma_S S(t) dW^Q_{S}(t), \nonumber \\
dr(t) &=& (\theta(t) - \alpha  r(t))  dt + \sigma_{r} dW^Q_{r}(t),
\nonumber
\end{eqnarray}

\noindent where the dynamics of the fund $S(t)$ and the interest
rates $r(t)$ are linked by the correlation structure:

\begin{eqnarray}
E^{Q}[dW^Q_{S} dW^Q_{r}] = \rho_{Sr}  dt, \nonumber \\
\end{eqnarray}

In the BSHW model  $\mu_x$  and $\sigma_x$ are given by

\begin{eqnarray}
 \mu_x &=& \frac{\rho_{Sr} \sigma_r \sigma_S}{\alpha} (1-e^{-\alpha T}),\nonumber \\
\sigma^2_x &=& \frac{ \sigma^2_r }{2\alpha} (1-e^{-2 \alpha T}).
\end{eqnarray}

\newpage

\section{Down-and-in GAO results \label{DIGAO_results}}

\begin{figure}[ht]
\begin{center}
\includegraphics[height=3.5cm]{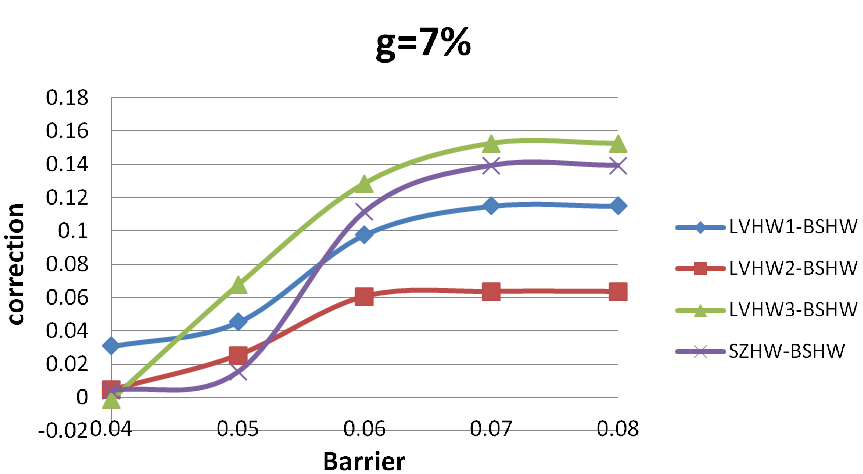}
\includegraphics[height=3.5cm]{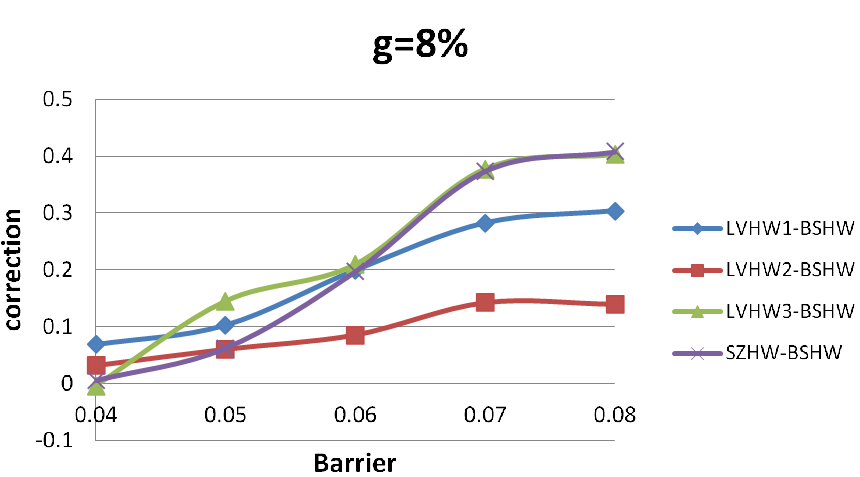}
\includegraphics[height=3.5cm]{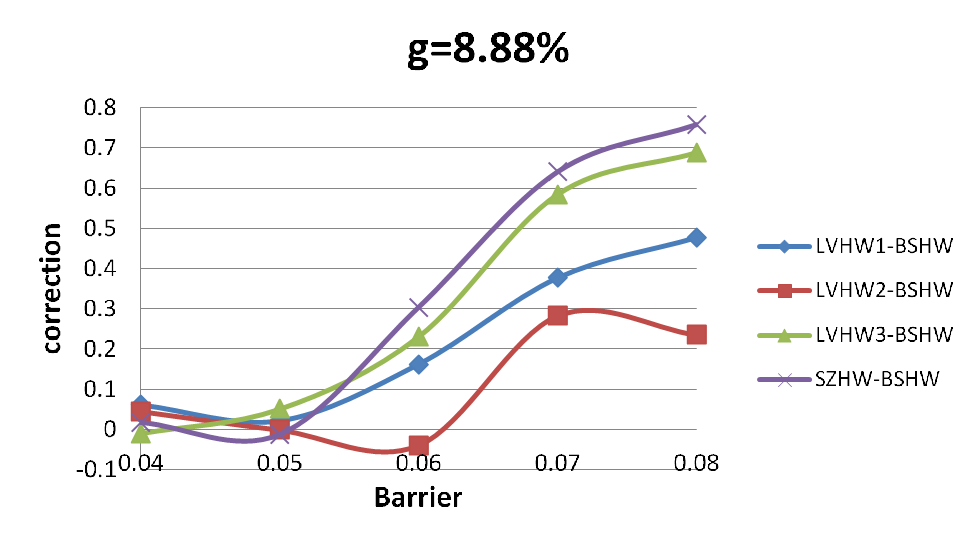}
\includegraphics[height=3.5cm]{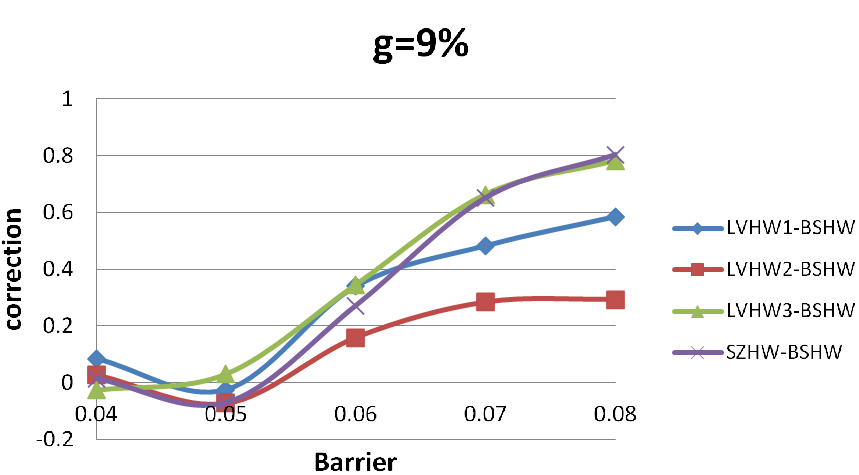}
\includegraphics[height=3.5cm]{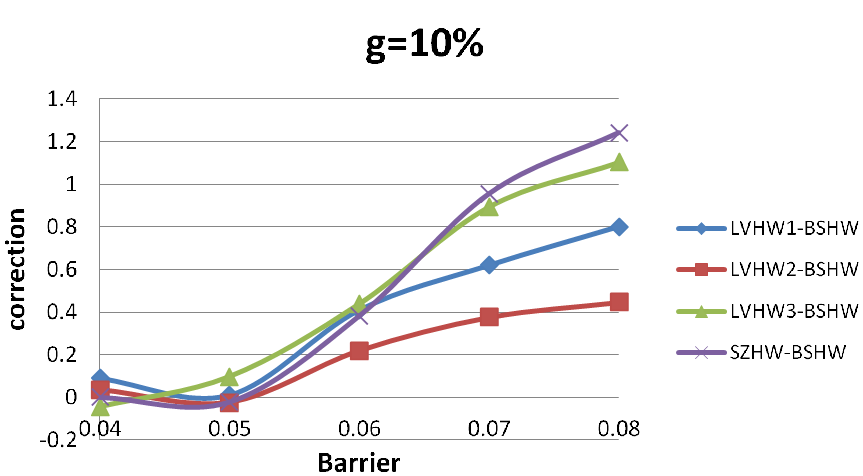}
\includegraphics[height=3.5cm]{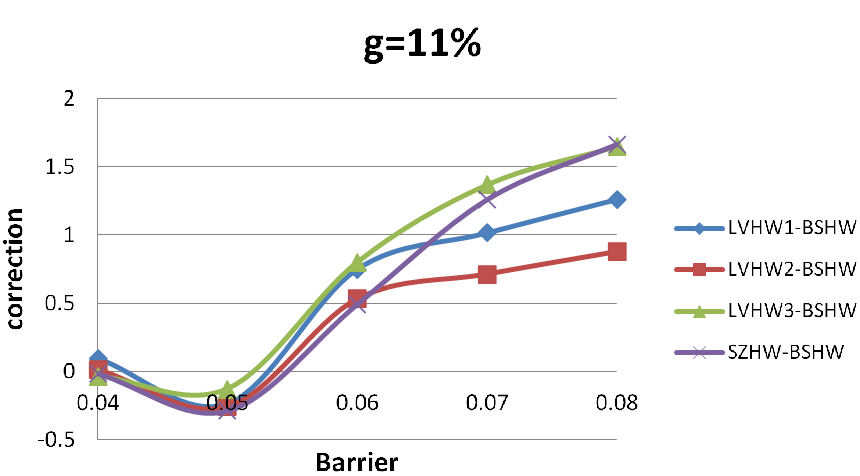}
\includegraphics[height=3.5cm]{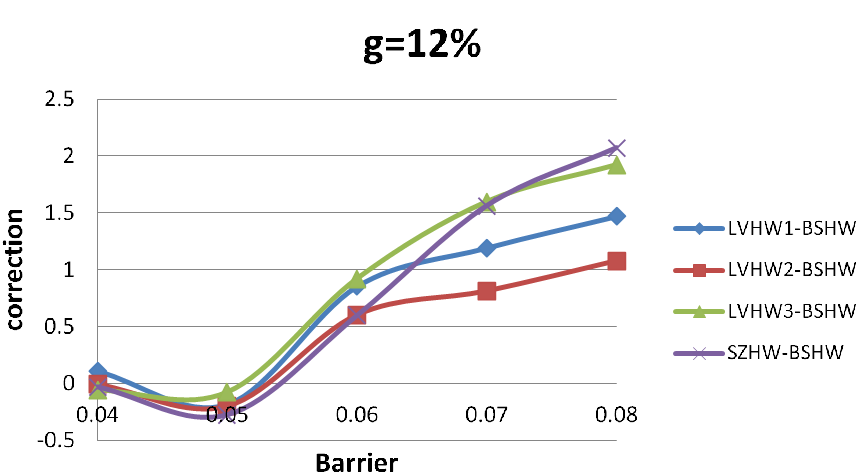}
\includegraphics[height=3.5cm]{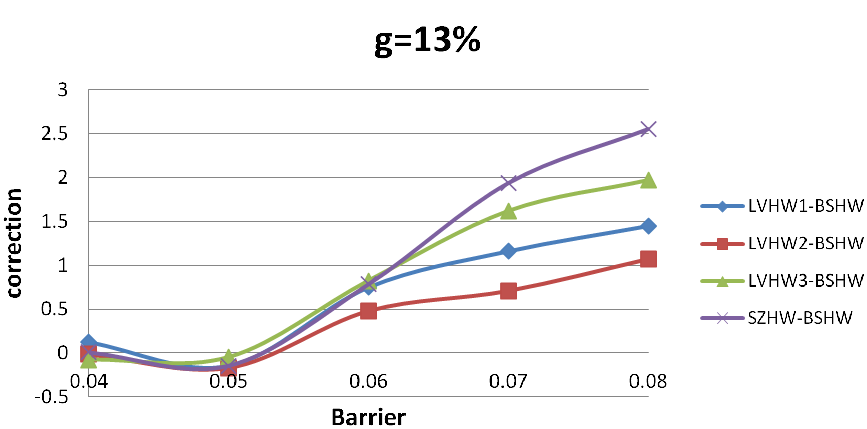}
\caption{Graphical representation of DI GAO corrections of the
SZHW, the LVHW1, LVHW2 and LVHW3 models for different guaranteed
annuity rates $g$ with respect to the BSHW model.}\label{fig:DI
GAO_corrections}
\end{center}
\end{figure}

\begin{table}\footnotesize
\begin{center}{
\begin{tabular}{|c|c|c|c|c|c|c|}
\hline
                     \multicolumn{ 6}{|c|}{{\bf BSHW DI GAO Total Value (US)  }} & {\bf BSHW GAO } \\
\hline
    \backslashbox{$g$}{Barrier} &       8.00\% &      7.00\% &       6.00\% &       5.00\% &       4.00\% &            \\
\hline
    7\% & \phantom{0}0.9069 & \phantom{0}0.9069  & \phantom{0}0.6754 & \phantom{0}0.0931 & \phantom{0}0.0023  & \phantom{0}0.9069  \\

      8\%& \phantom{0}3.1600 & \phantom{0}2.8189 & \phantom{0}1.3813 & \phantom{0}0.1053 & \phantom{0}0.0086 & \phantom{0}3.1600 \\

    8.88\%& \phantom{0}6.9412&  \phantom{0}5.0746 & \phantom{0}2.0215&  \phantom{0}0.2096& \phantom{0}0.0198 & \phantom{0}7.1019 \\

      9\%& \phantom{0}7.5111 & \phantom{0}5.3759 & \phantom{0}2.0906 & \phantom{0}0.2878&  \phantom{0}0.0385 & \phantom{0}7.7384 \\

       10\%&  13.3689 & \phantom{0}8.2064 & \phantom{0}2.8525 & \phantom{0}0.2979 & \phantom{0}0.0581 & 14.8802 \\

      11\%& 19.6122& 10.8281 & \phantom{0}3.3454 & \phantom{0}0.5995 & \phantom{0}0.0829 & 23.6438 \\

      12\%& 26.3137& 13.6888& \phantom{0}4.0752  &\phantom{0}0.6122& \phantom{0}0.1110 & 33.6896 \\

      13\% & 33.2845& 16.7626 & \phantom{0}5.0114 & \phantom{0}0.6437&  \phantom{0}0.1297&  44.3822 \\
\hline
                     \multicolumn{ 6}{|c|}{{\bf SZHW DI GAO Total Value (US)  }} & {\bf SZHW GAO } \\
\hline
    7\%& \phantom{0}1.0460&  \phantom{0}1.0460&  \phantom{0}0.7869&  \phantom{0}0.1085 & \phantom{0}0.0064 & \phantom{0}1.0460 \\

      8\% & \phantom{0}3.5678 & \phantom{0}3.1925& \phantom{0}1.5783  & \phantom{0}0.1680  & \phantom{0}0.0145 & \phantom{0}3.5678 \\

    8.88\% & \phantom{0}7.6988  & \phantom{0}5.7147 & \phantom{0}2.3250 & \phantom{0}0.1969 & \phantom{0}0.0372 & \phantom{0}7.8692 \\

      9\% & \phantom{0}8.3141 & \phantom{0}6.0276&  \phantom{0}2.3623 & \phantom{0}0.2168&  \phantom{0}0.0518 & \phantom{0}8.5554 \\

       10\%  & 14.6116& \phantom{0}9.1621&  \phantom{0}3.2343 & \phantom{0}0.2759&  \phantom{0}0.0586&  16.1414 \\

      11\% & 21.2745& 12.0856 & \phantom{0}3.8351 & \phantom{0}0.3093 & \phantom{0}0.0679 & 25.2674 \\

      12\% & 28.3855 &15.2518 & \phantom{0}4.6729 & \phantom{0}0.3367 & \phantom{0}0.0811  &35.5863 \\

      13\% & 35.8365 &18.6996& \phantom{0}5.7971 & \phantom{0}0.5005&  \phantom{0}0.1373 & 46.5705 \\
\hline
                     \multicolumn{ 6}{|c|}{{\bf LVHW1 DI GAO Total Value (US)  }} & {\bf LVHW1 GAO } \\
\hline
7\% &    \phantom{0}1.0218 & \phantom{0}1.0218 & \phantom{0}0.7731 & \phantom{0}0.1383  & \phantom{0}0.0331 & \phantom{0}1.0218 \\

      8\% & \phantom{0}3.4634 & \phantom{0}3.1009& \phantom{0}1.5804 & \phantom{0}0.2079  & \phantom{0}0.0772&  \phantom{0}3.4634 \\

  8.88\% &  \phantom{0}7.4179 & \phantom{0}5.4514 & \phantom{0}2.1841 & \phantom{0}0.2320 & \phantom{0}0.0817 & \phantom{0}7.5845\\

      9\% &   \phantom{0}8.0950 & \phantom{0}5.8578 & \phantom{0}2.4303 & \phantom{0}0.2630&  \phantom{0}0.1223 & \phantom{0}8.3261 \\

       10\% &   14.1694& \phantom{0}8.8259  & \phantom{0}3.2620 & \phantom{0}0.3096  & \phantom{0}0.1485  &15.6900 \\

      11\% &   20.8737 & 11.8444& \phantom{0}4.0955&  \phantom{0}0.3627 & \phantom{0}0.1801&  24.9009 \\

      12\% &   27.7819 &    14.8798 &     \phantom{0}4.9278 &     \phantom{0}0.4283 &      \phantom{0}0.2190 &    35.1173 \\

      13\% &    34.7312 &    17.9215 &     \phantom{0}5.7595 &     \phantom{0}0.4941 &     \phantom{0}0.2492 &    45.7425 \\
 \hline
 \multicolumn{ 6}{|c|}{{\bf LVHW2 DI GAO Total Value (US)  }} & {\bf LVHW2 GAO } \\
\hline
    7\% &    \phantom{0}0.9705 & \phantom{0}0.9705 & \phantom{0}0.7361 & \phantom{0}0.1184 & \phantom{0}0.0068 & \phantom{0}0.9705 \\

      8\% & \phantom{0}3.2996 & \phantom{0}2.9612 & \phantom{0}1.4667 & \phantom{0}0.1657 & \phantom{0}0.0401 & \phantom{0}3.2996 \\

    8.88\% &  \phantom{0}7.1769  & \phantom{0}5.3578 & \phantom{0}1.9822  & \phantom{0}0.2087 & \phantom{0}0.0651 & \phantom{0}7.3323 \\

      9\% &  \phantom{0}7.8041 & \phantom{0}5.6591&  \phantom{0}2.2500 & \phantom{0}0.2157  & \phantom{0}0.0657 & \phantom{0}8.0248 \\

       10\% &   \phantom{0}13.8150 & \phantom{0}8.5819 & \phantom{0}3.0712&  \phantom{0}0.2736 & \phantom{0}0.0936&  15.2712 \\

      11\% &   20.4883 &11.5410& \phantom{0}3.8766 & \phantom{0}0.3429 & \phantom{0}0.0996 & 24.3749 \\

      12\% &  27.3924 &14.5022 & \phantom{0}4.6793  & \phantom{0}0.4150& \phantom{0}0.1048  &34.5070  \\

      13\% &    34.3552 &17.4711 & \phantom{0}5.4916 & \phantom{0}0.4765 & \phantom{0}0.1133 & 45.0702 \\

   \hline
                     \multicolumn{ 6}{|c|}{{\bf LVHW3 DI GAO Total Value (US)  }} & {\bf LVHW3 GAO } \\
\hline
   7\% &    \phantom{0}1.0593 & \phantom{0}1.0593&  \phantom{0}0.8039 & \phantom{0}0.1605 & \phantom{0}0.0008 & \phantom{0}1.0593 \\

      8\% &    \phantom{0}3.5629 & \phantom{0}3.1964 & \phantom{0}1.5906&  \phantom{0}0.2499 & \phantom{0}0.0045 & \phantom{0}3.5629 \\

    8.88\% &  \phantom{0}7.6300 & \phantom{0}5.6593 & \phantom{0}2.2524  & \phantom{0}0.2608 & \phantom{0}0.0104  & \phantom{0}7.8085 \\

      9\% &   \phantom{0}8.2923 & \phantom{0}6.0394&  \phantom{0}2.4369 & \phantom{0}0.3192  & \phantom{0}0.0131 & \phantom{0}8.5334 \\

       10\% &  14.4721& \phantom{0}9.1002 & \phantom{0}3.2908 & \phantom{0}0.3953 & \phantom{0}0.0140 & 16.0349\\

      11\% &   21.2630 &12.1972& \phantom{0}4.1453  & \phantom{0}0.4700 & \phantom{0}0.0487&  25.3506 \\

      12\% &   28.2351& 15.2863 & \phantom{0}4.9954  & \phantom{0}0.5330& \phantom{0}0.0497&  35.6268 \\

      13\% &    35.2509 &18.3785 & \phantom{0}5.8336  & \phantom{0}0.5942 & \phantom{0}0.0516  &46.2961 \\
 \hline
\end{tabular}
\caption{Comparison of ``down-and-in GAO" total values given by
the BSHW and the SZHW and the LVHW models for eight different
guaranteed annuity rates $g$ and for five different barriers. The
pure GAO values for the eight different guaranteed annuity rates
$g$ are also given in the last column. \label{tab1:DI GAO total
value}}}
\end{center}
\end{table}

\newpage

\section{Graphics \label{appendix_graphics}}


\begin{figure}[h!]
\centering
\subfigure[implied volatility surface]{
\includegraphics[height=4.5cm]{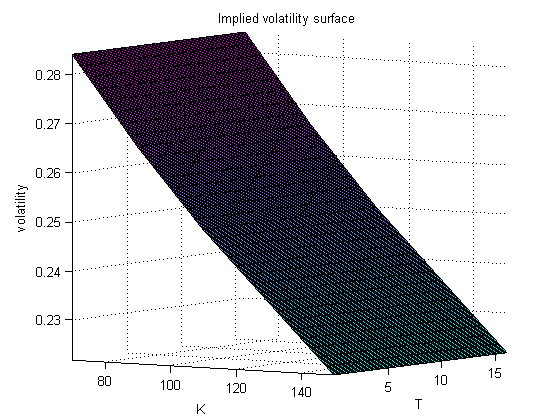}
\label{fig:US_impvoll_dsigdtnul} }
\subfigure[market call prices]{
\includegraphics[height=4.5cm]{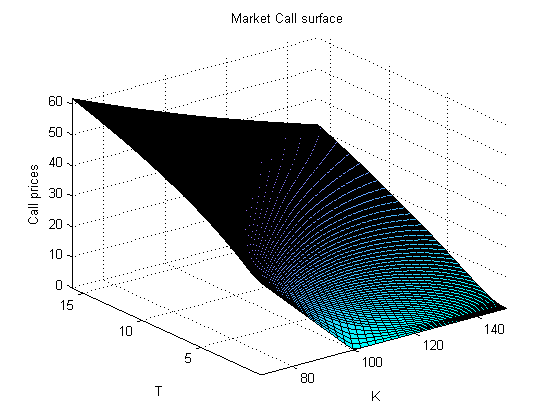}
\label{fig:US_mkt_call_dsigdtnul} }
\subfigure[local volatility (stochastic interest rates)]{
\includegraphics[height=4.5cm]{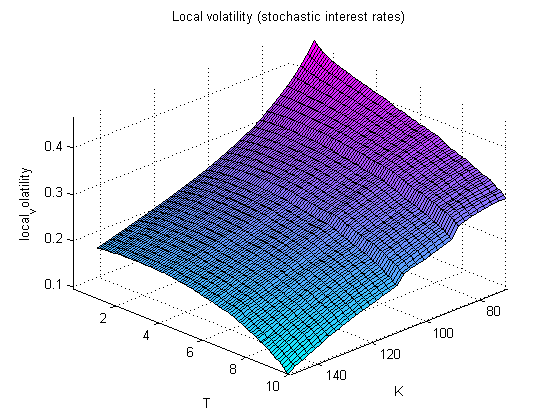}
\label{fig:US_locvol_fig_stoch_dsigdtnul} }
\subfigure[local volatility (constant interest rates)]{
\includegraphics[height=4.5cm]{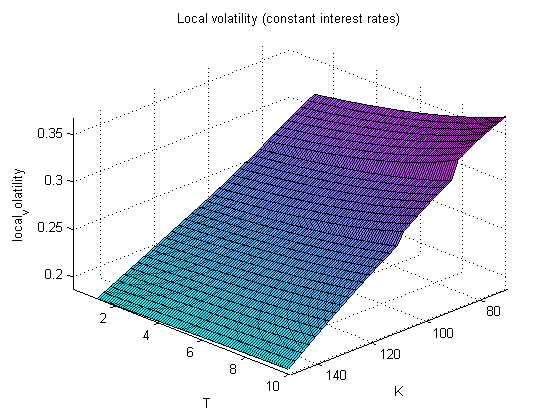}
\label{fig:US_locvol_fig_cst_dsigdtnul} }
\subfigure[difference]{
\includegraphics[height=4.5cm]{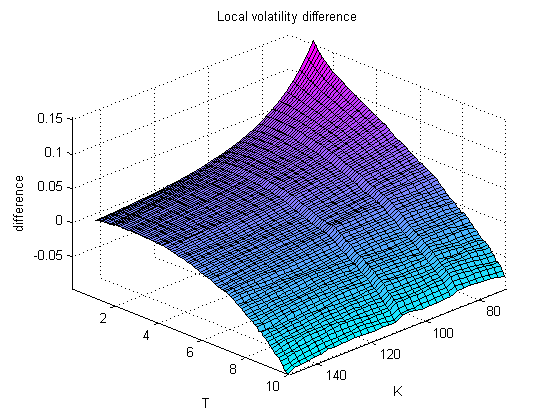}
\label{fig:US_locvol_fig_diff_dsigdtnul} }
\label{fig:US_dsigdtnul}
\caption[Optional caption for list of figures]{Plot of the implied
volatility (built from the 10-years S\&P500 implied volatility
curve and assuming a constant volatility with respect to the
maturity ($\frac{\partial \sigma_{imp}}{\partial T}=0.00$)), the
corresponding market call prices surface, the associated local
volatility obtained in both stochastic and constant interest rates
framework and finally, the difference between these two local
volatilities.}
\end{figure}

\begin{figure}[ht]
\centering
\subfigure[implied volatility surface]{
\includegraphics[height=5cm]{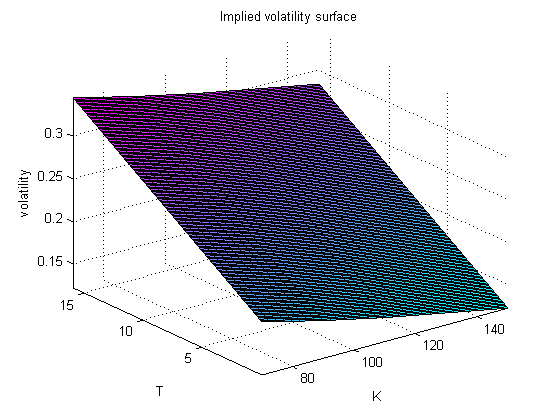}
\label{fig:US_impvoll_dsigdtpositif} }
\subfigure[market call prices]{
\includegraphics[height=5cm]{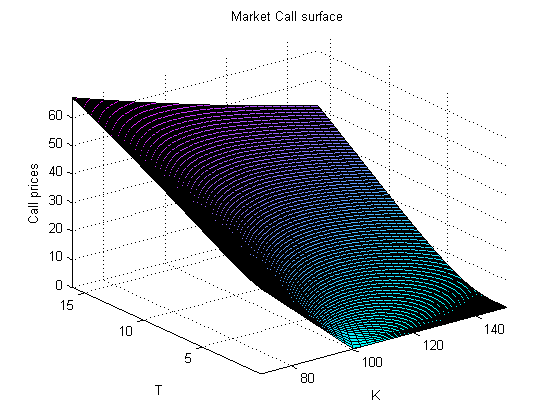}
\label{fig:US_mkt_call_dsigdtpositif} }
\subfigure[local volatility (stochastic interest rates)]{
\includegraphics[height=5cm]{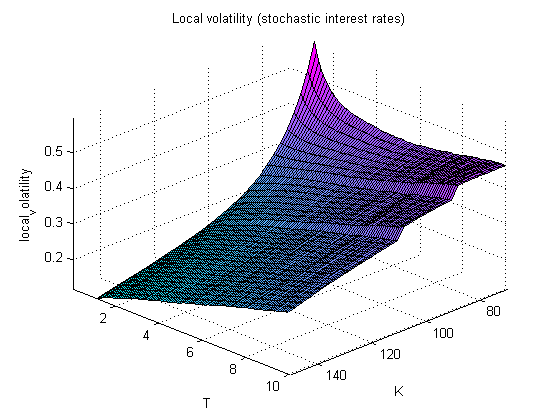}
\label{fig:US_locvol_fig_stoch_dsigdtpositif} }
\subfigure[local volatility (constant interest rates)]{
\includegraphics[height=5cm]{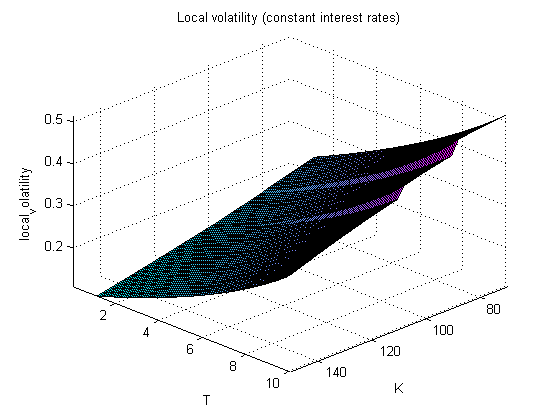}
\label{fig:US_locvol_fig_cst_dsigdtpositif} }
\subfigure[difference]{
\includegraphics[height=5cm]{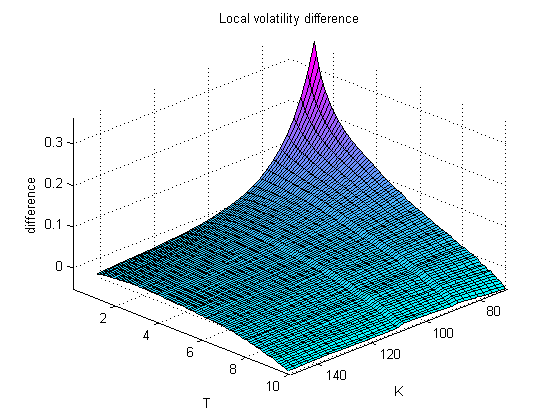}
\label{fig:US_locvol_fig_diff_dsigdtpositif} }
\label{fig:US_dsigdtpositif}
\caption[Optional caption for list of figures]{Plot of the implied
volatility (built from the 10-years S\&P500 implied volatility
curve and assuming an increasing volatility with respect to the
maturity ($\frac{\partial \sigma_{imp}}{\partial T}=0.01$)), the
corresponding market call prices surface, the associated local
volatility obtained in both stochastic and constant interest rates
framework and finally, the difference between these two local
volatilities.}
\end{figure}

\begin{figure}[ht]
\centering
\subfigure[implied volatility surface]{
\includegraphics[height=5cm]{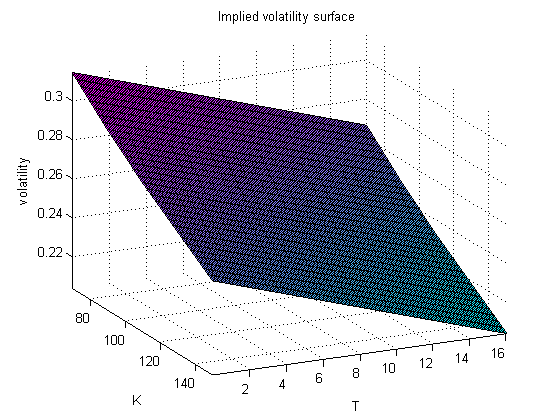}
\label{fig:US_impvoll_dsigdtnegatif} }
\subfigure[market call prices]{
\includegraphics[height=5cm]{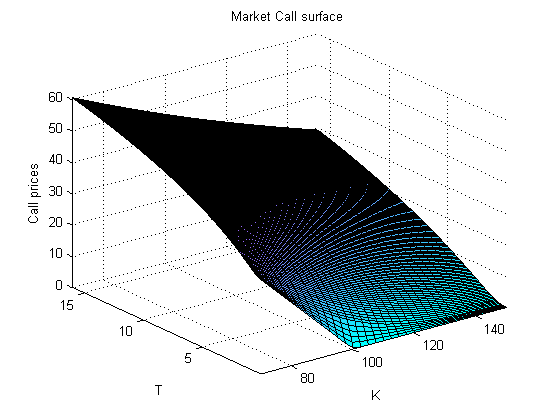}
\label{fig:US_mkt_call_dsigdtnegatif} }
\subfigure[local volatility (stochastic interest rates)]{
\includegraphics[height=5cm]{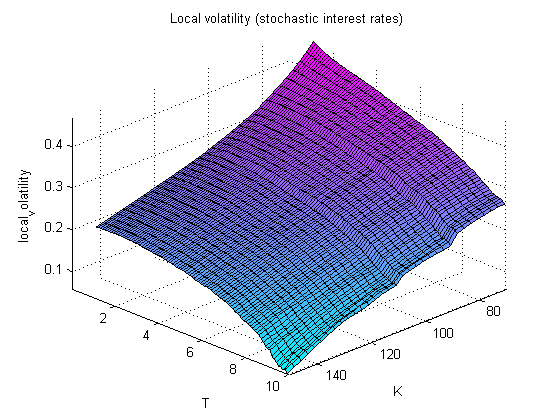}
\label{fig:US_locvol_fig_stoch_dsigdtnegatif} }
\subfigure[local volatility (constant interest rates)]{
\includegraphics[height=5cm]{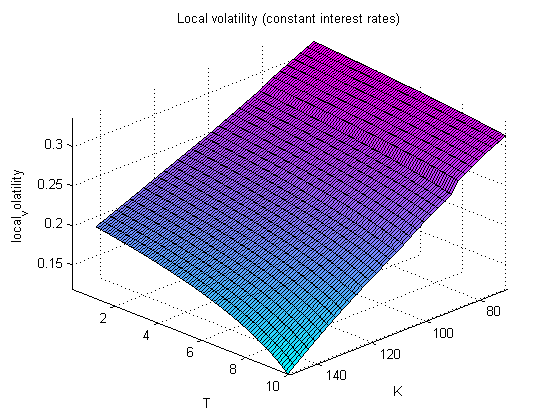}
\label{fig:US_locvol_fig_cst_dsigdtnegatif} }
\subfigure[difference]{
\includegraphics[height=5cm]{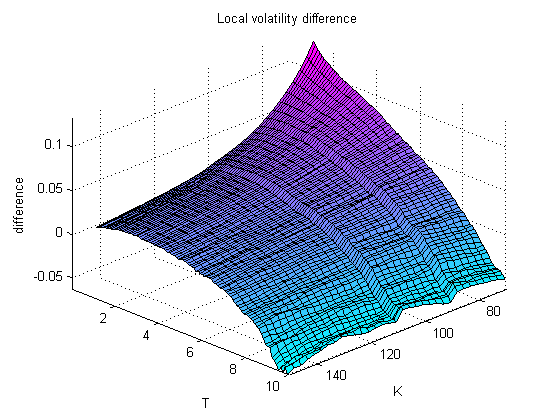}
\label{fig:US_locvol_fig_diff_dsigdtnegatif} }
\label{fig:US_dsigdtnegatif}
\caption[Optional caption for list of figures]{Plot of the implied
volatility (built from the 10-years S\&P500 implied volatility
curve and assuming an decreasing volatility with respect to the
maturity ($\frac{\partial \sigma_{imp}}{\partial T}=-0.003$)), the
corresponding market call prices surface, the associated local
volatility obtained in both stochastic and constant interest rates
framework and finally, the difference between these two local
volatilities.}
\end{figure}


\begin{figure}[ht]
\centering
\subfigure[implied volatility surface]{
\includegraphics[height=5cm]{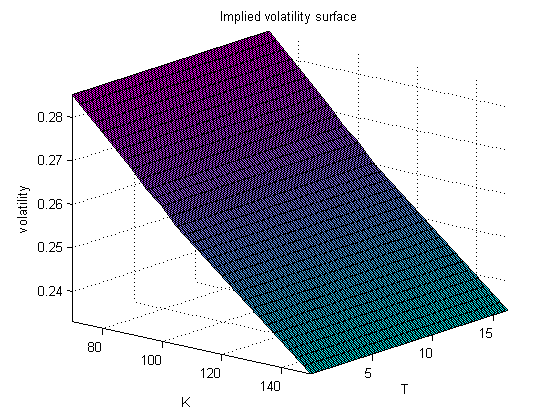}
\label{fig:EUR_impvoll_dsigdtnul} }
\subfigure[market call prices]{
\includegraphics[height=5cm]{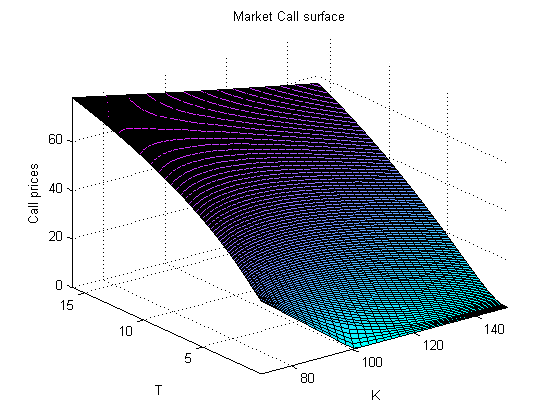}
\label{fig:EUR_mkt_call_dsigdtnul} }
\subfigure[local volatility (stochastic interest rates)]{
\includegraphics[height=5cm]{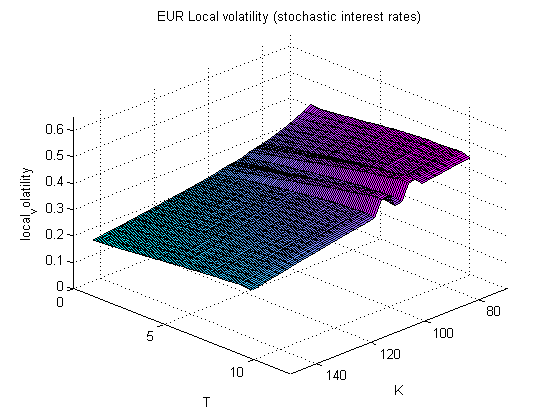}
\label{fig:EUR_locvol_fig_stoch_dsigdtnul} }
\subfigure[local volatility (constant interest rates)]{
\includegraphics[height=5cm]{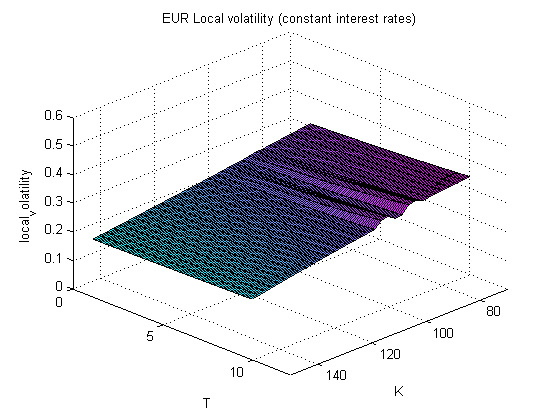}
\label{fig:EUR_locvol_fig_cst_dsigdtnul} }
\subfigure[difference]{
\includegraphics[height=5cm]{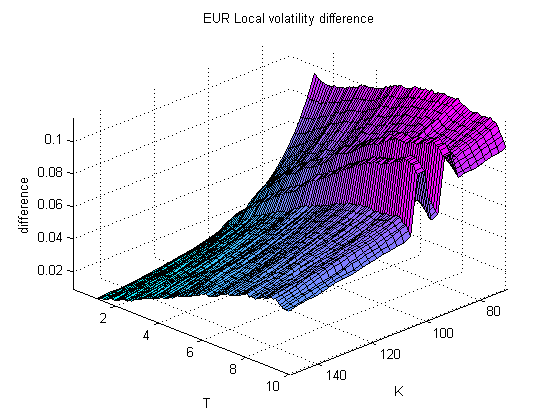}
\label{fig:EUR_locvol_fig_diff_dsigdtnul} }
\label{fig:EUR_dsigdtnul}
\caption[Optional caption for list of figures]{Plot of the implied
volatility (built from the 10-years EuroStoxx50 implied volatility
curve and assuming a constant volatility with respect to the
maturity ($\frac{\partial \sigma_{imp}}{\partial T}=0.00$)), the
corresponding market call prices surface, the associated local
volatility obtained in both stochastic and constant interest rates
framework and finally, the difference between these two local
volatilities. }
\end{figure}

\begin{figure}[ht]
\centering
\subfigure[implied volatility surface]{
\includegraphics[height=5cm]{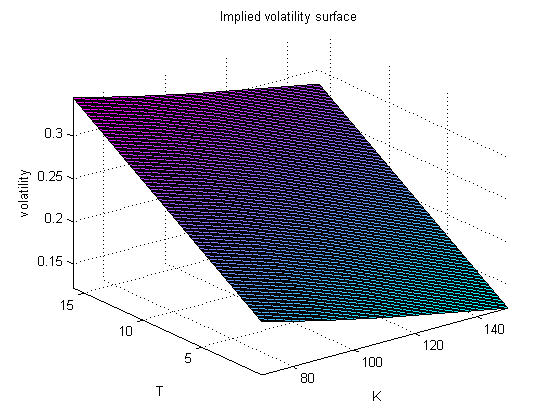}
\label{fig:EUR_impvoll_dsigdtpositif} }
\subfigure[market call prices]{
\includegraphics[height=5cm]{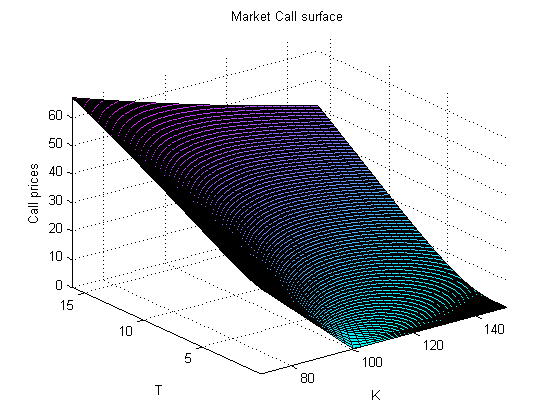}
\label{fig:EUR_mkt_call_dsigdtpositif} }
\subfigure[local volatility (stochastic interest rates)]{
\includegraphics[height=5cm]{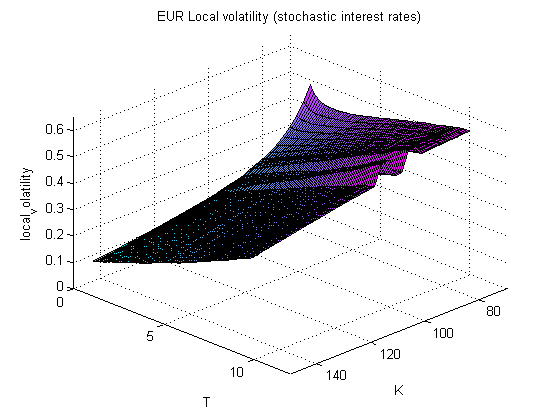}
\label{fig:EUR_locvol_fig_stoch_dsigdtpositif} }
\subfigure[local volatility (constant interest rates)]{
\includegraphics[height=5cm]{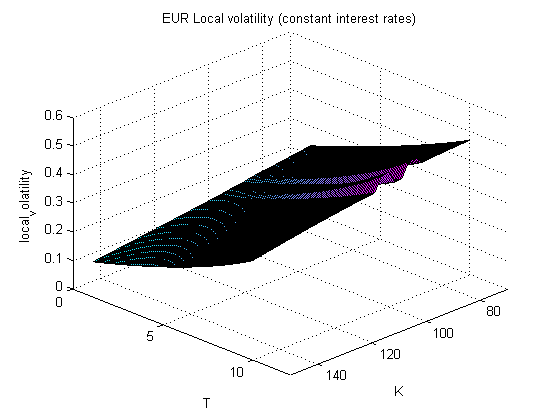}
\label{fig:EUR_locvol_fig_cst_dsigdtpositif} }
\subfigure[difference]{
\includegraphics[height=5cm]{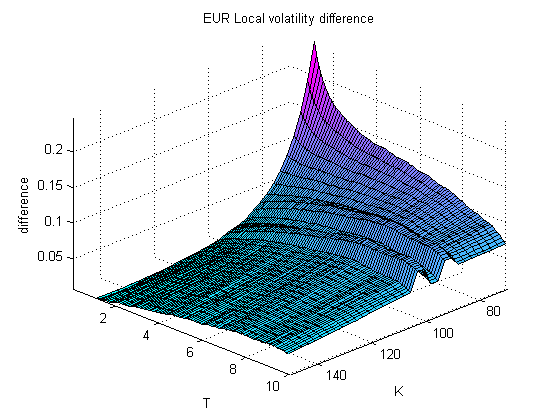}
\label{fig:EUR_locvol_fig_diff_dsigdtpositif} }
\label{fig:EUR_dsigdtpositif}
\caption[Optional caption for list of figures]{Plot of the implied
volatility (built from the 10-years EuroStoxx50 implied volatility
curve and assuming an increasing volatility with respect to the
maturity ($\frac{\partial \sigma_{imp}}{\partial T}=0.01$)), the
corresponding market call prices surface, the associated local
volatility obtained in both stochastic and constant interest rates
framework and finally, the difference between these two local
volatilities. }
\end{figure}

\begin{figure}[ht]
\centering
\subfigure[implied volatility surface]{
\includegraphics[height=5cm]{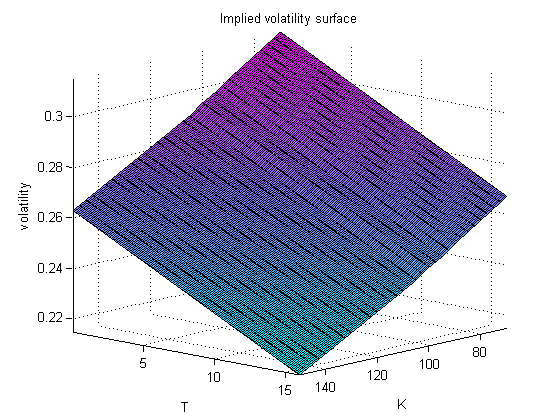}
\label{fig:EUR_impvoll_dsigdtnegatif} }
\subfigure[market call prices]{
\includegraphics[height=5cm]{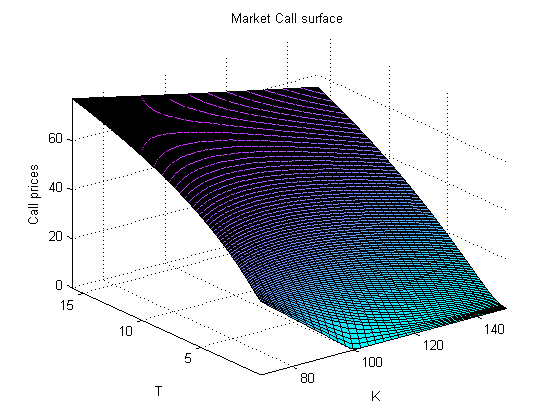}
\label{fig:EUR_mkt_call_dsigdtnegatif} }
\subfigure[local volatility (stochastic interest rates)]{
\includegraphics[height=5cm]{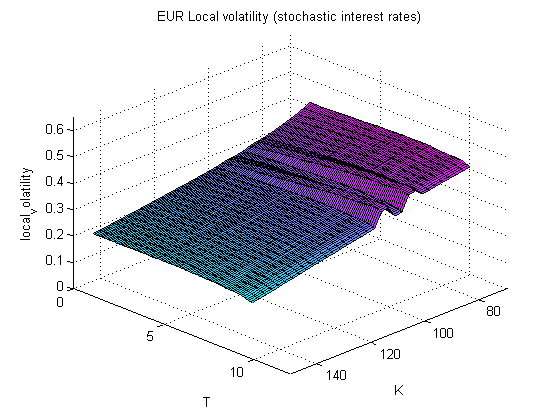}
\label{fig:EUR_locvol_fig_stoch_dsigdtnegatif} }
\subfigure[local volatility (constant interest rates)]{
\includegraphics[height=5cm]{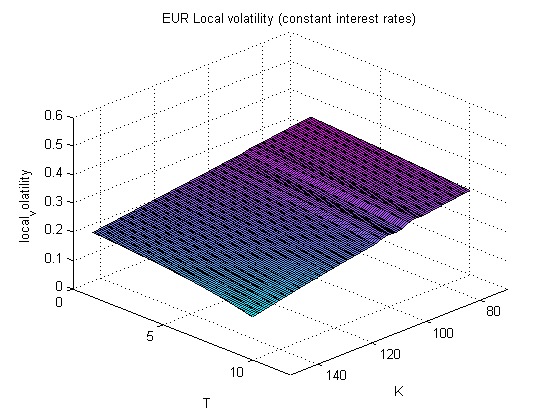}
\label{fig:EUR_locvol_fig_cst_dsigdtnegatif} }
\subfigure[difference]{
\includegraphics[height=5cm]{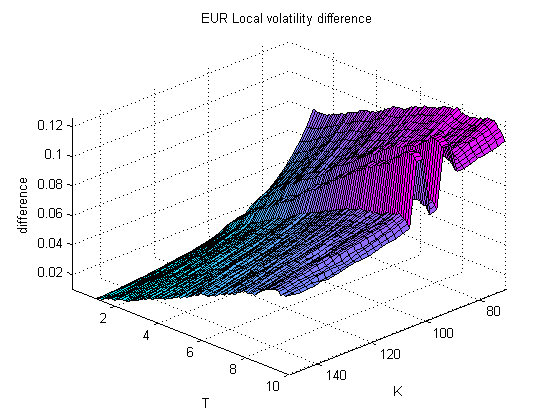}
\label{fig:EUR_locvol_fig_diff_dsigdtnegatif} }
\label{fig:EUR_dsigdtnegatif}
\caption[Optional caption for list of figures]{Plot of the implied
volatility (built from the 10-years EuroStoxx50 implied volatility
curve and assuming an decreasing volatility with respect to the
maturity ($\frac{\partial \sigma_{imp}}{\partial T}=-0.003$)), the
corresponding market call prices surface, the associated local
volatility obtained in both stochastic and constant interest rates
framework and finally, the difference between these two local
volatilities.}
\end{figure}

\end{document}